\documentclass[final,5p,times,twocolumn,authoryear]{elsarticle}

\usepackage{float}
\usepackage{listings}
\usepackage{booktabs}
\usepackage{amsmath}
\usepackage{array}
\usepackage{subfig, bm}

\usepackage{wrapfig}
\usepackage{enumitem}
\usepackage{tikz}
\usepackage{amssymb}
\usepackage{amsthm}
\usepackage{mdwlist}
\usepackage{mleftright}
\usepackage{hyperref}
\hypersetup{colorlinks,linkcolor={blue},citecolor={blue}} 
\usepackage[nameinlink]{cleveref} 
\usepackage{graphicx}
\usepackage[font=small,labelfont=bf]{caption}
\usepackage{subfiles}
\usepackage{import}
\usepackage{dsfont}
\usepackage{mathtools}
\usepackage{algorithm}
\usepackage{algorithmic}
\usepackage{sjmacros}
\usepackage{thmstyles}
\usepackage{stfloats}
\usepackage{cuted}
\usepackage{xcolor}

\journal{Automatica}

\begin{document}
\let\WriteBookmarks\relax
\def\floatpagepagefraction{1}
\def\textpagefraction{.001}
\allowdisplaybreaks

\begin{frontmatter}
    \title{
    Predictive Control of Linear Discrete-Time Markovian \\Jump Systems \authorrevised{by Learning Recurrent Patterns}
    }
    
    \author[a]{SooJean Han$^*$}
    \affiliation[a]{organization={Department of Computing and Mathematical Sciences},
    addressline={California Institute of Technology},
    city={Pasadena},
    postcode={91125}, 
    state={CA},
    country={USA}}
    \fnmark[*]
    \cortext[1]{Corresponding author (e-mail: \href{mailto:soojean@caltech.edu}{soojean@caltech.edu})}
    \fntext[1]{This work is supported by the National Science Foundation Graduate Research Fellowship under Grant No. DGE‐1745301 and the Aerospace Corporation.
    }
    
    \author[a]{Soon-Jo Chung}
    
    \author[a]{John C. Doyle}
\end{frontmatter}

\begin{abstract}
    Incorporating \textit{pattern-learning for prediction (PLP)} in many discrete-time or discrete-event systems allows for computation-efficient controller 
    \authorrevised{
    design by memorizing patterns to schedule control policies based on their future occurrences.
    }
    In this paper, we demonstrate the effect of PLP by designing a controller architecture for a class of linear Markovian jump systems (MJS) where the aforementioned ``patterns'' correspond to finite-length sequences of modes.
    In our analysis of recurrent patterns, we use martingale theory to derive closed-form solutions to quantities pertaining to the occurrence of patterns: 1) the expected minimum occurrence time of any pattern from some predefined collection, 2) the probability of a pattern being the first to occur among the collection.
    Our method is applicable to real-world dynamics because we make two extensions to common assumptions in prior pattern-occurrence literature.
    First, the distribution of the mode process is unknown, and second, the true realization of the mode process is not observable.
    As demonstration, we consider fault-tolerant control of a dynamic topology-switching network, and empirically compare PLP to 
    \authorrevised{
    two controllers without PLP: a baseline based on the novel System Level Synthesis (SLS) approach and a topology-robust extension of the SLS baseline.
    }
    We show that PLP is able to reject disturbances as effectively as the topology-robust controller at reduced computation time and control effort.
    We discuss several important tradeoffs, such as the size of the pattern collection and the system scale versus the accuracy of the mode predictions, which show how different PLP implementations affect stabilization 
    \authorrevised{and runtime}
    performance.
\end{abstract}

\begin{highlights}
\setlength\itemsep{0pt}
\item By considering linear discrete-time Markovian jump systems (MJS) with unknown mode-switching dynamics, we make concrete the broad notion that controller synthesis can be made more efficient by reducing computation time and redundancy via Pattern-Learning for Prediction (PLP), which learns patterns in the underlying mode process, stores them into memory, and predicts their future occurrences.

\item The PLP component of the controller architecture leverages martingale methods from prior literature, but with two important extensions that make it more suitable for real-world MJS applications: 1) the distribution of the mode process is unknown, and 2) the realization of the mode process over time is not observable.

\item We apply our proposed architecture to fault-tolerant control of a network with dynamic topology, and perform an extensive numerical study which compares the performance of the PLP controller against a baseline and a topology-robust extension of the baseline.
Our study also provides insights into important tradeoffs that emphasize the impact of PLP, e.g., the size of the pattern collection and the system scale versus the accuracy of the mode predictions.
A controller with PLP is able to match the control effort of the baseline, maintain a disturbance-rejection error similar to the topology-robust controller, and achieve runtime faster than either.
\end{highlights}


\maketitle


\section{Introduction}
Model-based controller synthesis methods can be developed for stochastic systems if a theoretical characterization of their stochastic process distribution exists.
In the literature, this concept is most notable for Gaussian white noise systems~\citep{doyle78,reif99,theodorou10} or MJS~\citep{xiong05,shi15}. 
Our prior work~\cite{han21ijrnc} suggested the possibility of expanding such methods to Poisson shot noise perturbations.
For many discrete-time or discrete-event systems, we can take advantage of the fact that the underlying stochastic process is a sequence of random variables which occurs as repeated patterns of interest.
For example, in fault-tolerance control or manufacturing process applications, a pattern of interest may be a specific sequence of modes which corresponds to a critical system fault~\citep{cho98,hanmer13_book}.
Another example can be found in queuing-based systems such as vehicle intersection networks~\citep{boon16,leeuwaarden06}, where repetition arises naturally when counting entities in the queue over time.

\authorrevised{Learning pattern repetitions in the underlying stochastic process of many discrete-time or discrete-event stochastic systems allows for at least two ways of more efficient controller design.}
First, we may store past sample paths of the stochastic process into memory so that if a certain pattern occurs multiple times, we do not need to recompute the corresponding control policy at every occurrence.
Second, we may predict the expected occurrence times of patterns in the future and schedule to apply the corresponding control policies at the predicted times.
This idea is present in many applications.
For example, a collision-avoidance trajectory over a future horizon of time can be computed for a moving vehicle based on repeated experiences of obstacle behavior, instead of relying only on instantaneous measurements of each obstacle's position~\citep{richards06,mesquita12,shim12}.
In the class of discrete-event systems, labeled transition representations are invoked to solve fault diagnosis and prediction problems because they enable easier identification of repeated patterns over time~\citep{jeron08,jeron06}.

\vspace{-10pt}
\subsection{Related Work}
\textit{Reducing Repetitive Computation}: Making 
\authorrevised{
control more efficient in terms of computation time by taking advantage of any repetition in the system behavior}
is a fairly common concept in the engineering community.
For example,~\cite{chen17} proposes repetitive learning control for a class of nonlinear systems tracking reference signals that are periodic.
\cite{zheng21} discusses a method to approximate linear Gaussian systems using a hidden Markov model (HMM), then trains it by exploiting its periodic structures.
Some notable machine learning approaches for control, i.e. long short-term memory networks (LSTMs) and imitation learning~\citep{verma19}, are also designed to reduce redundant learning.
\authorrevised{
In fact, the broad class of meta-learning algorithms refers to algorithms which not only focus on learning the subject matter (e.g. classification tasks), but also on learning the learning procedure itself~\citep{oconnell22}.
For problems that can be solved using deep reinforcement learning methods, \textit{experience replay}~\citep{fedus20a} manages to improve sample and data efficiency by storing the last few experiences into memory and ``replaying'' them.
A related approach is called \textit{episodic control}~\citep{lengyel07,blundell16,pritzel17}, which incorporates \textit{episodic memory}~\citep{botvinick19} into traditional learning techniques with the goal of speeding up training by recalling specific instances of highly rewarding experiences.
Towards this end, numerous episodic control approaches have been proposed, including model-free episodic control~\cite{blundell16} and neural episodic control~\cite{pritzel17}.
In our paper~\cite{han23l4dc}, we consider vehicle traffic congestion control over urban networks of signalized intersections where the controller architecture leverages an extension of episodic control which uses equivalence classes to limit the growth of the memory table.
}

%
\textit{Computing Pattern-Occurrence Quantities}: Repetition in stochastic processes can be addressed in theory by solving ``pattern-occurrence problems'', which characterizes one or more specific sequences of values as ``patterns'' then solves for quantities such as the expected time until their next observation in the stochastic process.
Scan statistics~\citep{pozdnyakov14} is a popular tool founded on martingale theory, and is often used to characterize the distribution of pattern occurrences in applications dealing such as fault-tolerance and anomaly-detection.
For example,~\cite{guerriero09} uses a scan statistics approach for distributed target-sensing using stationary sensors and a moving agent under simplified assumptions on the distribution of the sensors' positions.
Formulas for predicting the occurrence of patterns have been derived when the patterns emerge from an i.i.d. sequence (see, e.g.,~\cite{shuo_yen_li80},~\cite{gerber_li_81}, and~\cite{pozdnyakov06}) and when the patterns are generated from scalar Markov chains~\citep{glaz06,pozdnyakov08}.

\textit{Controlling Uncertain Systems}: One notable drawback to current pattern-occurrence methods~\citep{pozdnyakov06,glaz06,pozdnyakov08} are their reliance on the assumptions that we are able to precisely observe the stochastic process and that its distribution is known.
\authorrevised{
In fact, there is an abundance of research in system identification and data-driven control, e.g.,~\cite{dean18safely} and~\cite{ho21a}, because these assumptions often do not hold in real world applications.
\cite{ho21a} considers robust and adaptive control for nonlinear systems with large model uncertainties by using a nested convex body chasing approach to optimally choose an approximate model around which the control law is designed.
Many of these algorithms involve a natural multi-step procedure where the original uncertain dynamics and constraints are mapped down to an approximate model, which is then used for planning and control.
For example,~\cite{nakka21} first develops a surrogate optimization problem with chance constraints by leveraging polynomial chaos expansion before generating approximate solution trajectories via sequential convex programming.
}

\authorrevised{
\textit{Predictions for Structured Control}:
Using the memorized previous patterns and state/control trajectories, some algorithms in the literature have also invoked predictions to reduce redundant computation.
\textit{Model predictive control (MPC)}~\citep{garcia89mpc,cuzzola02} is one of the most popular methodologies that demonstrates this, and both short-term and long-term predictions for online control have been proven to be beneficial even in the face of either purely stochastic or adversarial disturbances~\citep{chen15wierman}.
In~\cite{yu20wierman}, this is demonstrated explicitly by applying greedy conventional MPC to the linear quadratic tracking problem, and proving near-optimality in the dynamic regret performance metric.
\cite{nagabandi17} provides an architecture which combines learning with MPC for robot link manipulation tasks; MPC is used for control law design based on the dynamics of the robotic arm approximated through learning and additional data is used only if the performance of the current model falls short of the desired goal, making the entire procedure efficient in time and data consumption.
MPC has also been developed for specific classes of systems; in particular,~\cite{park02} considers MPC for discrete-time MJS when the dynamics are linear and uncertain.
The benefit of predictions is especially notable when there is spatial or temporal structure to the problem.
\textit{Graph neural networks (GNNs)}~\cite{battaglia18} are an example of a learning-based approach which encodes the topology of the graph for tasks such as graph classification and representation learning~\cite{kipf17}.
Recently, extensions of GNNs are also being used for congestion control problems in computer networks~\cite{rusek20} and vehicle traffic forecasting~\cite{roseYu18,cui18}; both applications deal with large-scale networks for which exploitable spatial and temporal repetitions are abundant.
}

\vspace{-5pt}
\subsection{Contributions}
\authorrevised{This paper aims to demonstrate the effectiveness and benefit of \textit{Pattern-Learning for Prediction (PLP)} on controller synthesis for a class of linear MJS whose underlying mode-switching dynamics are unknown.
In this context}, ``patterns'' are recurrent finite-length sequences of modes that arise in the MJS; 
\authorrevised{
PLP uses these patterns to make control design more efficient by memorizing certain patterns to prevent the re-computation of the control laws associated with them, then scheduling control laws for patterns that may occur in the future.
}
Our architecture \authorrevised{for the class of uncertain linear discrete-time MJS} consists of three components.
First, \authorrevised{\textit{Mode Process Identification (ID)}} uses state and control sequences to learn the unknown statistics of the mode process; \authorrevised{here}, these are the transition probability matrix (TPM) and the mode at the current time.
Second, \textit{\authorrevised{PLP}} uses the estimated TPM and current mode to compute quantities pertaining to the future occurrence of patterns.
\authorrevised{Third, \textit{Control Law Design}} performs the appropriate optimization to compute the control law associated with each pattern when it first occurs.

\authorrevised{We develop and integrate the PLP} component in an otherwise straightforward architecture which leverages well-researched techniques in system identification and predictive control.
In our analysis of recurring patterns, we use martingale theory to derive mathematical expressions for two quantities pertaining to the prediction of patterns: the expected minimum occurrence time of any pattern from some (user-defined) collection of patterns, and the probability of a pattern being the first to occur among the collection at the expected \authorrevised{time}.
Our method operates on two key extensions of prior pattern-occurrence literature (e.g.,~\cite{glaz06},~\cite{pozdnyakov14}) which makes it applicable to real-world dynamics: the distribution of the mode process is unknown, and the mode process over time is not observable \authorrevised{(e.g., the past and current modes the system has been in is unknown)}.
To our knowledge, our proposed architecture is the first to apply a martingale method to the learning-based control of a stochastic system.

We provide an extensive comparison study that demonstrates the effects of PLP on a version of the proposed three-part architecture applied to the control of a network with dynamic topology, where the modes correspond to the different possible topology variations.
\authorrevised{
For the purposes of this application, the controller architecture integrates two additional algorithms from existing literature.
First, MPC is used to schedule future control policies to be applied at the occurrence times specified by PLP.
Second, the novel system level synthesis (SLS) approach~\cite{wang18,anderson19} formulates the actual optimization problem to be solved; we especially use the data-driven formulation~\citep{xue21,carmen22} because of the uncertainties in the system.
The comparison is performed against two controllers based on SLS: a baseline SLS controller and an extension of SLS that was explicitly designed for topology robustness~\citep{han20l4dc}.
}
Our results offer insights into several important tradeoffs among four performance metrics which determine how PLP affects a controller's performance in stabilizing the system.
\authorrevised{
Compared to the baseline controller, we show that a PLP controller is able to achieve better disturbance-rejection at significantly reduced computation time and redundancy.
Furthermore, because Pattern-Learning can be viewed as an additional mode estimation algorithm for a suitable collection of patterns, it enables the estimated mode to match the true mode more often than without Pattern-Learning, boosting system identification performance.
We show that PLP can reject disturbances as well as the topology-robust controller while consuming less computation time and control effort, then discuss the role of system scale on PLP design criteria such as the choice of pattern collection.}
\authorrevised{
\subsection{Organization}
The rest of the paper is organized as follows.
In~\sec{setup}, we introduce the relevant notations, assumptions, and set up the uncertain linear discrete-time MJS considered throughout the entire paper.
\sec{framework} provides a coherent overview of the three components that make up our proposed controller architecture.
The subsequent sections go into further detail about the concrete choice of algorithms used to implement each component: Mode Process Identification (ID) in~\sec{mode_id}, PLP in~\sec{pattern_learning}, and (Predictive) Control Law Design in Sec.~\ref{sec:mode_control}.
We implement our controller architecture on a topology-changing network in~\sec{case_study}, and compare its performance against a couple of baseline controllers without PLP.
We conclude the paper in~\sec{conclusion}.
}
%
\bgroup
\def\arraystretch{1.5}
\begin{table}[t]
    \begin{center}
      {\footnotesize
      \begin{tabular}{| c | c |}
        \hline
        Sym. & Definition \\ \hline \hline
        \authorrevised{$\Delta T$} & Timescale of mode w.r.t. system (Assum.~\ref{assum:timescale})\\ \hline
        $\hat{\varphi}_{n}^{(t)}$ & Est. current mode at time $t$, $n\triangleq N[t]$ (Sec.~\ref{subsec:mode_id}) \\ \hline
        $\hat{P}^{(t)}[m_1,m_2]$ & Est. TPM entry for $m_1,m_2\in\Xcal$ (Sec.~\ref{subsec:mode_id}) \\ \hline
        $\Ccal[t]$ & Set of consistent modes at time $t$~\eqn{consistent_set} \\ \hline
        $\Psi[t]$ & Time-varying pattern collection (Defs.\ref{def:patterns},\ref{def:pattern_collection_tv}) \\ \hline
        $\boldsymbol{\psi}_k$ & A pattern from $\Psi$, enum. $k\in\{1,\cdots,K\}$ (Def.\ref{def:patterns}) \\ \hline
        $L$ & Future horizon, pattern length (Def.~\ref{def:pattern_collection_tv}) \\ \hline
        $\Ucal$ & Control law table in memory (Prop.~\ref{prop:mode_mpc_memory})\\ \hline
        $\hat{\tau}^{(t)}$ & Min. occurrence time of $\Psi[t]$ (Def.\ref{def:min_tau_probs}, Rmk.\ref{rmk:shift_indices}) \\ \hline
        $\hat{q}_k^{(t)}$ & First occurrence prob. of $\boldsymbol{\psi}_k\in\Psi[t]$ (Def.\ref{def:min_tau_probs}, Rmk.\ref{rmk:shift_indices}) \\ \hline
        $\Gamma$ & Augmented pattern collection~\eqn{augmented_collection} \\ \hline
        $\boldsymbol{\gamma}_{\ell}$ & Augmented pattern, enum. $\ell\leq\abs{\Xcal}^2\abs{\Psi}$ (Def.~\ref{def:augmented_patterns}) \\ \hline
        $\Scal_I^{(0)}$ & Set of Case 0 initial-ending strings (Def.\ref{def:ending_strings}) \\ \hline
        $\Scal_I^{(1)}$ & Set of Case 1 initial-ending strings (Def.\ref{def:ending_strings}) \\ \hline
        $\Scal_L$ & Set of later-ending strings (Def.\ref{def:ending_strings}) \\ \hline
        $\Scal$ & $=\Scal_L\cup\Scal_I = \Scal_L\cup(\cup_{i\in\{0,1\}}\Scal_I^{(i)})$ (Def.\ref{def:ending_strings}) \\ \hline
        $K_I^{(\chi)}$ & $=\abs{\Scal_I^{(\chi)}}, \chi\in\{0,1\}$ cardinality (Def.\ref{def:ending_strings}) \\ \hline
        $K_L$ & $=\abs{\Scal_L}$ cardinality (Def.\ref{def:ending_strings}) \\ \hline
        $\boldsymbol{\beta}_s$ & Ending string in $\Scal$, enum. $s\in\{1,\cdots,K_I+K_L\}$ \\ \hline
        $\Pbb(\boldsymbol{\beta}_s)$ & Prob. that $\boldsymbol{\beta}_s$ terminates $\{\xi_n\}$ (Def.~\ref{def:ending_string_prob}) \\ \hline
        $c_{\ell}$ & Initial reward of each type-$\ell$ agent (Def.\ref{def:agents_rewards}) \\ \hline
        $R_{\tau}^{(\ell)}$ & Type-$\ell$ cumu. net reward (Def.\ref{def:cum_net_rewards}) \\ \hline
        $R_{\tau}$ & Cumu. net reward (Def.\ref{def:cum_net_rewards}) \\ \hline
        $W_{s\ell}$ & Gain matrix entry $(s,\ell)$: total gain earned by\\
        & type-$\ell$ agent via ending string $\boldsymbol{\beta}_s$ (Def.\ref{def:gains_matrix_init_reward}) \\ \hline
      \end{tabular}
      }
    \end{center}
    \vspace{-10pt}
    \caption{Summary of \authorrevised{some of the notations used in the controller architecture}, listed in pairs of symbols (`Sym.') and definitions.
    \authorrevised{Many of these notations are used to develop the Pattern-Learning component (Sec.~\ref{sec:pattern_learning}).}
    }
    \vspace{-10pt}
    \label{tab:notation}
\end{table}
\egroup


\section{Setup and Preliminaries}\label{sec:setup}
We consider linear Markovian jump systems (MJS) of the following form:
\begin{align}\label{eq:plant_dynamics}
    \hskip1cm\xvect[t+1] &= A(\xi_{N[t]})\xvect[t] + B\uvect[t] + \wvect[t]
\end{align}
Here, $\xvect[t]{\,\in\,} \Rbb^{n_x}$ is the state, $A(\xi_{N[t]}){\,\in\,}\Rbb^{n_x\times n_x}$ is the dynamics matrix which changes according to the phase variable $\xi_{N[t]}$, $\uvect[t]{\,\in\,}\Rbb^{n_u}$ is the control input.
\authorrevised{
The external noise process $\wvect[t]{\,\in\,}\Rbb^{n_x}$ is unobservable and all we know about it is its upper norm bound $\norm{\wvect[t]}_{\infty}{\,\leq\,}\overline{w}$.
}
For each $t{\,\in\,}\Nbb$, $N[t]$ is the number of \textit{modes} (i.e., number of phase switches, or jumps arising from the underlying Markov chain) that have been observed by time $t$.
We say that the current \textit{mode-index} at time $t{\,\in\,}\Nbb$ is $n{\,\in\,}\Nbb$ if $N[t]{\,=\,}n$, and the transition from mode $\xi_{n-1}$ to $\xi_n$ occurs at time $T_n{\,\triangleq\,}\min\{s{\,\in\,}\Nbb\,|\,N[s] = n\}$.
The discrete mode process $\{\xi_n\}_{n=1}^{\infty}$ takes values from the set $\Xcal{\,\triangleq\,}\{1, \cdots, M\}$, where $M{\,\in\,}\Nbb$, and is defined such that $\xi_n{\,:\,}\Omega\to\Xcal$ on probability space $(\Omega, \Fcal, \Pbb)$ with filtration $\{\Fcal_n\}_{n=1}^{\infty}$, $\Fcal_n{\,\triangleq\,}\sigma(\xi_0, \xi_1, \cdots, \xi_n)$. 
\authorrevised{
We assume $B$ is a known constant matrix.
}

A summary of some of the most important notations used throughout the paper is provided in~\tab{notation}.
Throughout this paper, the letter $\xi$ is specifically reserved to denote random variable modes.
We distinguish $\{\xi_n\}$ from the sequence of deterministic values $\{\varphi_n\}$ which it takes, i.e., $\xi_n{\,=\,}\varphi_n$ for all past mode-indices $n{\,\in\,}\Nbb$.
Mode sequences denoted using other Greek letters are deterministic unless explicitly stated otherwise.
We henceforth denote all sequences of the form $\{\cdot\}_{n=1}^{\infty}$ using the shorthand notation $\{\cdot\}$, e.g., $\{\xi_n\}_{n=1}^{\infty}{\,\equiv\,}\{\xi_n\}$ and $\{\Fcal_n\}_{n=1}^{\infty}{\,\equiv\,}\{\Fcal_n\}$, and denote $\xvect[s{\,:\,}t]{\,=\,}\{\xvect[s], \cdots, \xvect[t]\}$ for any $s{\,<\,}t$, likewise for $\uvect[s{\,:\,}t],\wvect[s{\,:\,}t]$.
For any two $n,m{\,\in\,}\Nbb$ such that $n_1{\,<\,}n_2$, we denote random vectors of mode sequences $\xi_{n_1:n_2}{\,\triangleq\,} (\xi_{n_1}, \xi_{n_1+1}, \cdots, \xi_{n_2})$, and likewise $\varphi_{n_1:n_2}$.
We denote the concatenation of $\boldsymbol{\alpha}{\,\triangleq\,}(\alpha_1, \cdots, \alpha_a)$ and $\boldsymbol{\beta}{\,\triangleq\,}(\beta_1, \cdots, \beta_b)$ as $\boldsymbol{\alpha}\circ \boldsymbol{\beta}\triangleq(\alpha_1, \cdots, \alpha_a,\beta_1, \cdots, \beta_b)$, where $\boldsymbol{\alpha}$ and $\boldsymbol{\beta}$ are placeholders for either deterministic or random mode sequences.

\begin{assumption}\label{assum:timescale}
    The mode process $\{\xi_n\}$ operates on a timescale which is $\Delta T{\,\in\,}\Nbb$ times longer than the timescale of the system~\eqn{plant_dynamics}, i.e. if $N[t] = n$, then $N[t+a\Delta T] = n+a$ for any $a{\,\in\,}\Nbb$.
    This means $T_n - T_{n-1} = \Delta T$ for all $n{\,\in\,}\Nbb$.
    In certain applications, $\Delta T$ can be interpreted as the minimum time needed between switching modes, and for simplicity we assume that its value is known.
    Consequently, we assume that $N[t]$ and the sequence of transition times $\{T_n\}$ are also known. 
\end{assumption}

The mode process $\{\xi_n\}$ is generated from an irreducible Markov chain over the state-space $\Xcal$ with transition probability matrix (TPM) denoted by $P{\,\in\,}\Rbb^{M\times M}$ and initial probability vector $\pvect_0{\,\triangleq\,} [\text{p}_0(1),\cdots,\text{p}_0(M)]^\top{\,\in\,}\{0,1\}^M$.
We represent the entries of the TPM using brackets, so that $P[m_1, m_2]$ denotes the probability of the mode switching from $m_1$ to $m_2$, for any $m_1, m_2{\,\in\,}\Xcal$.
Suppose the probability distribution of $\xi_n$ is given by $\pvect_n{\,\in\,} [0,1]^M$ at mode-index $n{\,\in\,}\Nbb$. 
Then the mode process dynamics are updated in the usual Markov chain way $\pvect_{n+1}^\top{\,=\,}\pvect_n^\top P$.
This implies that given $\xi_n {\,=\,} \varphi_n{\,\in\,}\Xcal$, we have $\xi_{n+1}{\,=\,}m$ with probability $P[\varphi_n,m]$ for any $m\in\Xcal$.

\begin{assumption}\label{assum:knowns}
    \authorrevised{To demonstrate PLP and focus on the mode process, we take the simpler setting where the state $\xvect[t]$ is fully-observable; we thus design state-feedback control policies.
    Following the setup of bounded model errors in robust control theory, we assume $\overline{w}$ is known or otherwise attainable from small-gain theorems~\citep{zhou_doyle_book} or techniques based on structured singular values~\citep{doyle82}.
    For the mode process,} in addition to knowing the values of $\Delta T$, $N[t]$, and $\{T_n\}$ (see~\assum{timescale}), we consider the following settings.
    The true realizations $\{\varphi_n\}$ of the mode process $\{\xi_n\}$ are unknown over time, but the set $\Xcal$ of values that it takes and the initial mode $\xi_0{\,=\,}\varphi_0$ are known.
    The sparsity structure of the TPM $P$ is known, but the values of the nonzero entries are unknown.
\end{assumption}

\section{\authorrevised{Outline of} the Controller \authorrevised{Architecture}}\label{sec:framework}
The controller architecture we propose is visualized in~\fig{two_part_hierarchical_flow}.
It consists of 
\authorrevised{
three main parts: 1) Mode Process Identification (ID), 2) Pattern-Learning for Prediction (PLP) on the mode process, and 3) Control Law Design for the system dynamics.
In this section, we provide a brief description of each part--including an introduction of the main notations used--to provide a coherent view of the architecture (\fig{two_part_hierarchical_flow}) as a whole.
The details of each individual part and the choice of algorithms used to implement them are discussed in the subsequent sections: Mode Process ID in~\sec{mode_id}, PLP in~\sec{pattern_learning}, and Control Law Design in Sec.~\ref{sec:mode_control}.
We emphasize that our choice of algorithm to implement each component is unique to the uncertain linear discrete-time MJS setup described in~\sec{setup} and that alternative implementations can be made for other dynamics.
For example, in our paper~\cite{han23l4dc}, the controller architecture was designed for the specific application of vehicle traffic congestion control over urban networks of signalized intersections, in which the problem is set up as a Markov decision process.
}

\subsection{Mode Process Identification \authorrevised{Overview}}\label{subsec:mode_id}
\authorrevised{
For each time $t{\,\in\,}\Nbb$ and corresponding mode-index $n{\,\triangleq\,}N[t]$, the system maintains the following estimated statistics about the mode process $\{\xi_n\}$ and system dynamics~\eqn{plant_dynamics}: an estimate $\hat{P}^{(t)}$ of the true TPM $P$, and an estimate $\hat{\varphi}_{n}^{(t)}$ of the current mode $\varphi_{n}$.
The first part of our architecture, \textit{Mode Process Identification (ID)}, is responsible for learning these unknown statistics of the mode process.
Due to this uncertainty in the dynamics, we use hats and $(t)$ superscripts to emphasize that these quantities are estimates which change over time; as we will see in Sec.~\ref{sec:mode_id}, this is because modes are estimated based on state and control trajectories $\xvect[0{\,:\,}t], \uvect[0{\,:\,}t]$.
}

\begin{figure}
    \centering
    \includegraphics[width=0.95\columnwidth]{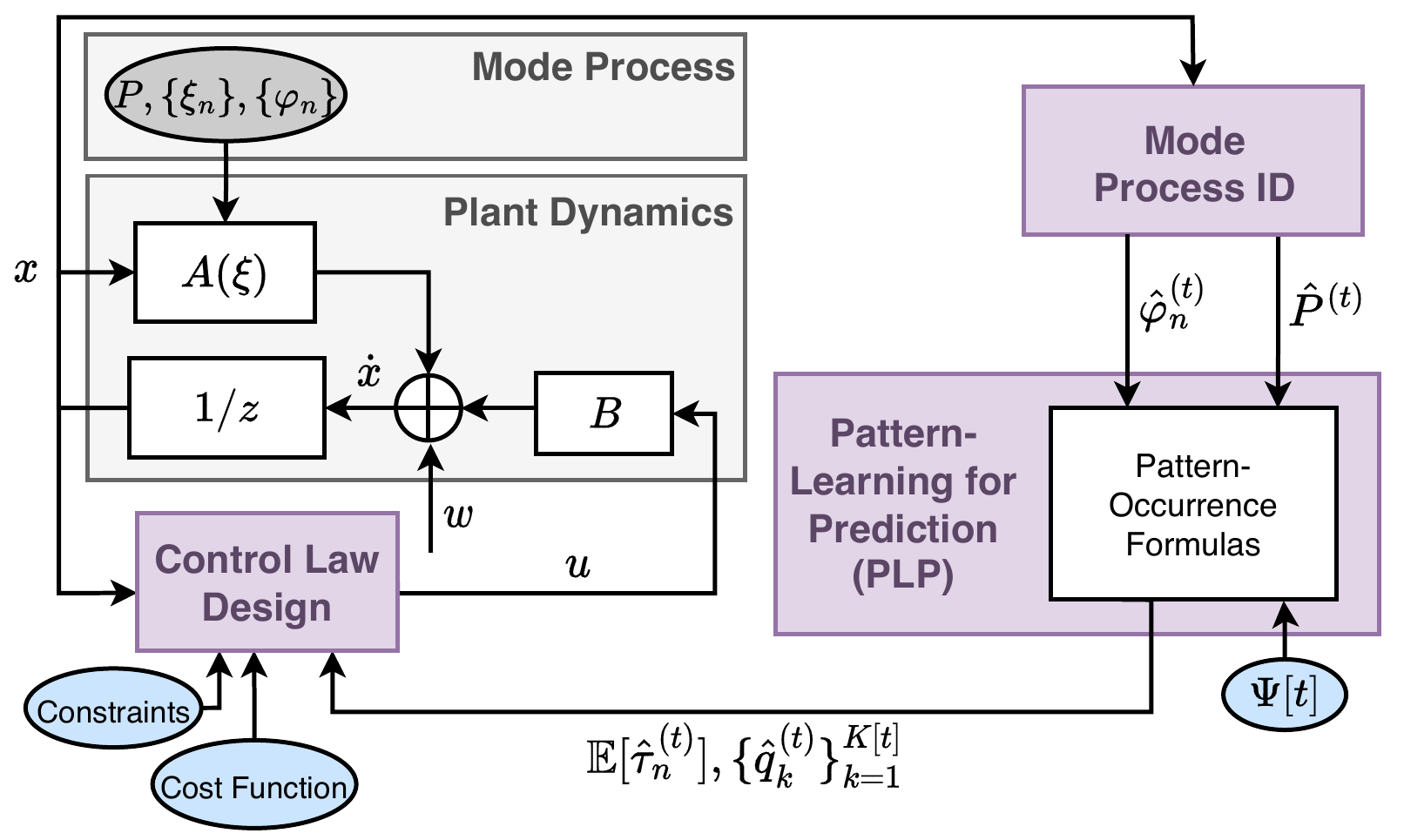}
    \caption{A flow diagram representation of the proposed controller architecture specifically for linear MJS dynamics of the form~\eqn{plant_dynamics}.
    Circles represent inputs to the algorithm; user-defined inputs are colored blue and unknown/unobservable parameters are colored gray.
    \authorrevised{
    The architecture consists of three main parts (violet boxes): 1) Mode Process ID (Sec.~\ref{subsec:mode_id}; Sec.~\ref{sec:mode_id}), 2) Pattern-Learning for Prediction (Sec.~\ref{subsec:pattern_learning}; Sec.~\ref{sec:pattern_learning}), 3) and Control Law Design (Sec.~\ref{subsec:mode_control}; Sec.~\ref{sec:mode_control}).
    }
    }
    \vspace{-5pt}
    \label{fig:two_part_hierarchical_flow}
\end{figure}

\authorrevised{
\subsection{\authorrevised{Setup of} Pattern-Learning for Prediction}\label{subsec:pattern_learning}
}
Once $\hat{P}^{(t)}$ and $\hat{\varphi}_{n}^{(t)}$ are obtained from  Mode Process ID (Sec.~\ref{subsec:mode_id}) for each $t{\,\in\,}\Nbb$ and $n{\,\triangleq\,}N[t]$, \textit{Pattern-Learning for Prediction (PLP)} in~\fig{two_part_hierarchical_flow}
\authorrevised{
computes additional statistics about the mode process (called the ``pattern-occurrence quantities'') that facilitate the creation of \textit{predictions}, which will be used in the Control Law Design component.
}

\authorrevised{
\begin{definition}[Prediction Horizon]\label{def:pred_horz}
    Define the constant $L{\,\in\,}\Nbb$ to be the \textit{prediction horizon} on the mode process, i.e., the length of the sequences of modes.
\end{definition}
}

In this paper, ``patterns'' refer to \authorrevised{length-$L$} sequences of modes in the mode process underlying the system~\eqn{plant_dynamics}, formalized in the following definition.

\begin{definition}[Patterns]\label{def:patterns}
    \authorrevised{
    Let $L{\,\in\,}\Nbb$ be the prediction horizon from~\defin{pred_horz}.}
    Define the set $\Psi \triangleq \{\boldsymbol{\psi}_1, \cdots, \boldsymbol{\psi}_K\}$ 
    \authorrevised{
    to be a \textit{collection of patterns},
    }
    where each $\boldsymbol{\psi}_k\triangleq(\psi_{k,1}, \cdots, \psi_{k,L})$ is a mode sequence with length $L$ and elements $\psi_{k,j}{\,\in\,}\Xcal$.
    Each $\boldsymbol{\psi}_k$ is referred to as a \textit{(mode) pattern} if we are interested in observing its occurrence in the mode process $\{\xi_n\}$ over time (e.g., because it models a system fault).
\end{definition}

\authorrevised{It is possible for the patterns in $\Psi$ to have different lengths, e.g., $\boldsymbol{\psi}_k\triangleq(\psi_{k,1}, \cdots, \psi_{k,d_k})$ for any $d_k{\,\in\,}\Nbb$.
However, in the context of predicting the future modes of an MJS like~\eqn{plant_dynamics}, it is probabilistically more likely to observe patterns with shorter lengths; for balance, we keep each pattern the same length $L$.
}

\begin{definition}\label{def:feasible}
    A pattern or an arbitrary sequence of modes $(\alpha_1, \cdots, \alpha_a)$ with length $a{\,\in\,}\Nbb$ is \textit{feasible \authorrevised{with respect to $\hat{P}^{(t)}$}} if it can be generated by the Markov chain with TPM $\hat{P}^{(t)}$, i.e., $\hat{P}^{(t)}[\alpha_i, \alpha_{i+1}]>0$ for all $i{\,\in\,}\{1,\cdots, a-1\}$.
\end{definition}

\authorrevised{
Because the statistics of the mode process are estimates instead of true values, it becomes necessary to consider a pattern collection $\Psi$ (from~\defin{patterns}) which varies with time.

\begin{definition}[Time-Varying Collection]\label{def:pattern_collection_tv}
    We construct the collection of patterns $\Psi[t]$, with time-varying cardinality $K[t]$, to be a \authorrevised{subset} of feasible length-$L$ future sequences of modes given the estimated current mode $\hat{\varphi}_{n}^{(t)}$:
    \begin{align}\label{eq:mpc_pattern_collection}
        \Psi[t]&\triangleq\{\boldsymbol{\psi}_1^{(t)}, \cdots, \boldsymbol{\psi}_{K[t]}^{(t)}\}\notag\\
        &\subseteq\{\text{feasible }(\alpha_1,\cdots,\alpha_L)| \hat{P}^{(t)}[\hat{\varphi}^{(t)}_{n}, \alpha_1]>0, \alpha_i{\,\in\,}\Xcal\}
    \end{align}
\end{definition}
}\vspace{-8pt}

\begin{definition}[Pattern-Occurrence Times]\label{def:pattern_times}
    Denote $n{\,\triangleq\,} N[t]\in\Nbb$ to be the current mode-index at current time $t{\,\in\,}\Nbb$, and suppose the estimated current mode is $\xi_{n}{\,=\,}\hat{\varphi}_{n}^{(t)}$.
    Then for each of the patterns in the collection $\Psi$ from~\defin{patterns}, define the following stopping times for each $k{\,\in\,}\{1,\cdots, K[t]\}$:
    \begin{align}\label{eq:pattern_def}
        &\hskip.3cm\hat{\tau}_{k|n}^{(t)}\!\triangleq\!\min\{i\in\Nbb\,|\,\xi_{n} = \hat{\varphi}_{n}^{(t)},\xi_{n+i-L+1:n+i}\!\!=\! \boldsymbol{\psi}_k^{(t)}\}
    \end{align}
\end{definition}

\begin{definition}[Time and Probability of First Occurrence]\label{def:min_tau_probs}
    Under the setup of~\defin{pattern_times} suppose $\xi_{n+\hat{\tau}_{n}^{(t)}-L+1:n+\hat{\tau}_{n}^{(t)}}=\boldsymbol{\psi}_k^{(t)}$.
    Then define the following for the collection $\Psi$: 
    \begin{align}\label{eq:min_tau}
        \hskip.5cm\hat{\tau}_{n}^{(t)}\!\triangleq\!\! \min_{k\in\{1,\cdots,K[t]\}}\!\!\hat{\tau}_{k|n}^{(t)},\hskip.5cm \hat{q}_k^{(t)} \!\triangleq\! \Pbb(\hat{\tau}_{n}^{(t)}\!=\!\hat{\tau}_{k|n}^{(t)})
    \end{align}
\end{definition}

\begin{problem}[Pattern-Occurrence Quantities]\label{prob:pattern_occurrence}
    \authorrevised{To generate predictions from the mode process}, we are interested in characterizing the following \textit{pattern-occurrence quantities} described in~\defin{min_tau_probs}.
    \begin{itemize}[leftmargin=*]
        \setlength\itemsep{0em}
        \item the estimate $\Ebb[\hat{\tau}_{n}^{(t)}]$ of the \textit{mean minimum occurrence time}, which counts the number of mode-indices to observe the occurrence of any pattern from $\Psi[t]$, given the estimated current mode $\hat{\varphi}_{n}^{(t)}$.
        \item the estimated \textit{first-occurrence probabilities} $\{\hat{q}_k^{(t)}\}_{k=1}^{K[t]}$, where $\hat{q}_k^{(t)}{\,\in\,}[0,1]$ is the probability that pattern $\boldsymbol{\psi}_k{\,\in\,}\Psi[t]$ is the first to be observed among all of $\Psi[t]$.
    \end{itemize}
\end{problem}

Again, we keep the hat and superscript $(t)$ in the $\tau$ and $q_k$ quantities because we emphasize they are dependent on $\hat{P}^{(t)}$ and $\hat{\varphi}_{n}^{(t)}$ from Sec.~\ref{subsec:mode_id}, which may change over time.

\subsection{Predictive Control Law Design: General Formulation}\label{subsec:mode_control}
\authorrevised{
Let $g{\,:\,}\Rbb^{+}\times\Xcal\times\Rbb^{n_x}\to\Rbb^{n_u}$ be a generic function representing the mode-dependent state-feedback control law designed by the Control Law Design component in~\fig{two_part_hierarchical_flow}.
The Control Law Design component uses the expected occurrence time $\Ebb[\hat{\tau}_n^{(t)}]$ and probabilities $\{\hat{q}_k\}_{k=1}^{K[t]}$ computed from PLP (Sec.~\ref{subsec:pattern_learning}) to store the control policies of previously-occurred patterns and to schedule control policies in advance.
This procedure is described more carefully in the following two propositions.
}

\begin{proposition}[Scheduling Future Control Inputs]\label{prop:mode_mpc_schedule}
    Suppose we are given the estimated pattern-occurrence quantities $\Ebb[\hat{\tau}_{n}^{(t)}]$ and $\{\hat{q}_k^{(t)}\}_k$ from PLP.
    Let $\tau{\,\equiv\,}\Ebb[\hat{\tau}_{n}^{(t)}]$ be the shorthand notation (with a temporary abuse of notation) for the estimated expected minimum occurrence time for the specific pattern collection $\Psi[t]$ given estimated current mode $\hat{\varphi}_{n}^{(t)}$.
    To schedule a control law in advance, we simply choose the pattern $\boldsymbol{\psi}_k^{(t)}{\,\in\,}\Psi[t]$ corresponding to the largest occurrence probability $\hat{q}_k^{(t)}$.
    Then, until mode-index $\tau$, the future sequence of control inputs $\uvect[t{\,:\,}T_{n+\tau+1}-1]$ is
    \vspace{-15pt}
    \authorrevised{
    \begin{align}\label{eq:pattern_control}
        \hskip1cm\uvect[s] &= g(s,\psi_{k,1}^{(t)},\xvect[s]),\hskip.1cm s\in[t{\,:\,}T_{n+1}-1]\\
        &\vdots\notag\\
        \uvect[s] &= g(s,\psi_{k,L}^{(t)},\xvect[s]),\hskip.1cm s\in[T_{n+\floor{\tau}}{\,:\,}T_{n+\tau+1}-1]\notag
    \end{align}}
\end{proposition}
Aside from operating on a longer timescale (mode process instead of system dynamics), Proposition~\ref{prop:mode_mpc_schedule} is similar in principle to standard model predictive control (MPC): only the first control law in the sequence~\eqn{pattern_control}, corresponding to the first mode $\psi_{k,1}^{(t)}$, is applied at the next mode-index $n{\,+\,}\floor{\tau}$.

\begin{proposition}[Storing Past Control Inputs in Memory]\label{prop:mode_mpc_memory}
    Define $\Ucal$ to be a table which maps mode patterns $\boldsymbol{\psi}_k^{(t)}$ to \authorrevised{
    control policies $\{g(t,\psi_{k,1}^{(t)},\cdot), \cdots, g(t,\psi_{k,L}^{(t)},\cdot)\}$ 
    }
    and the accumulated state and control trajectories over each occurrence time.
    When $\boldsymbol{\psi}_k^{(t)}{\,\in\,}\Psi[t]$ is first observed, a new entry $\Ucal[\boldsymbol{\psi}_k^{(t)}](t)$, defined by~\eqn{pattern_control} for the specific $\boldsymbol{\psi}_k^{(t)}$, is created.
    For anticipated future occurrences of $\boldsymbol{\psi}_k^{(t)}$, the system schedules control inputs using $\Ucal[\boldsymbol{\psi}_k^{(t)}](t)$ in the form of~\eqn{pattern_control}.
    The entry for $\boldsymbol{\psi}_k^{(t)}$ is then updated at every occurrence time after its first.
\end{proposition}

\begin{figure}
    \centering
    \includegraphics[width=\columnwidth]{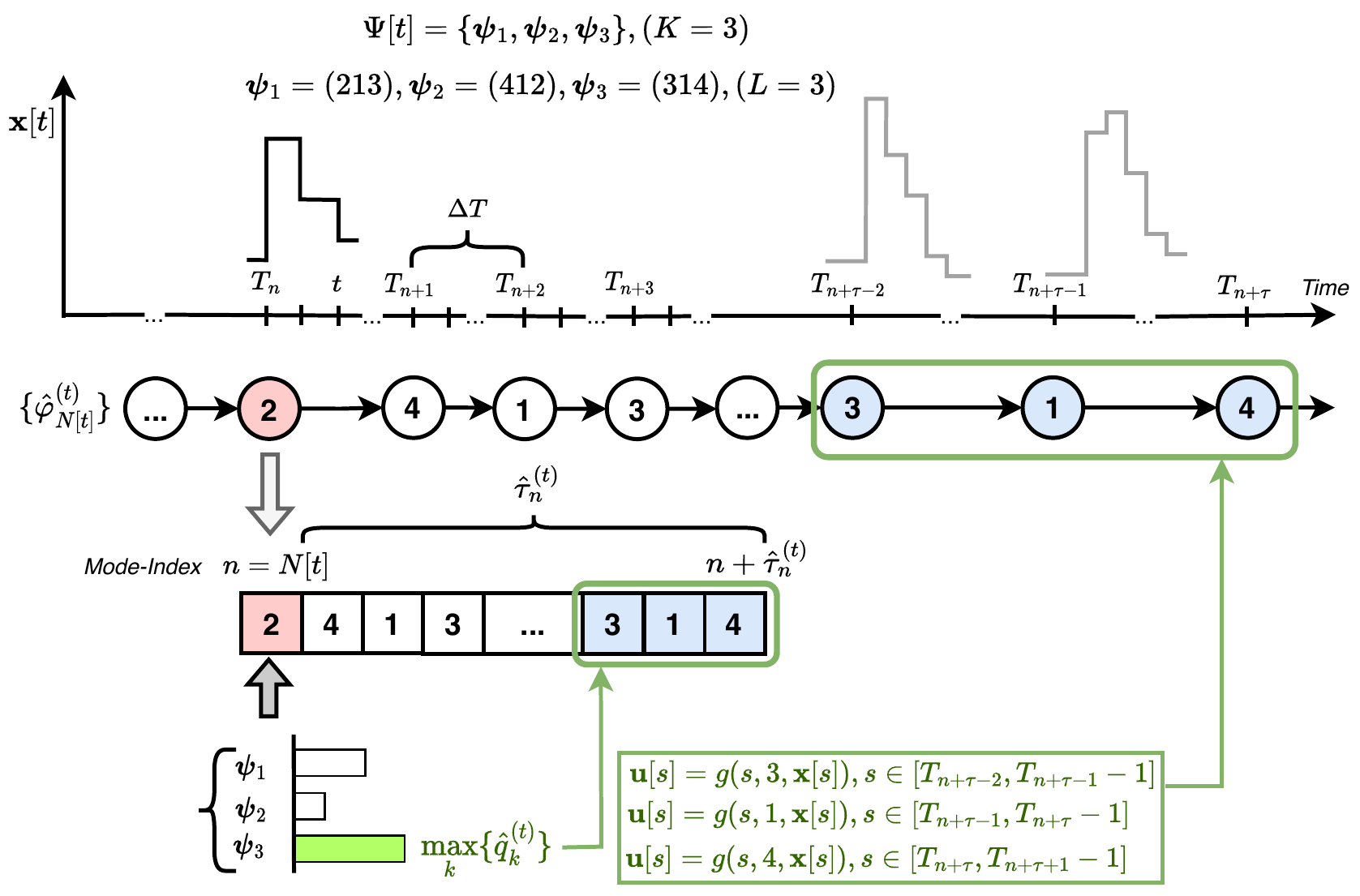}
    \caption{
    A visualization of PLP with a pattern collection of three patterns with $L=3$.
    The middle row of circles shows the evolution of $\hat{\varphi}_{n}^{(t)}$ over time, while the bottom row of boxes shows the process on the mode timescale (see~\assum{timescale}).
    The red circle and box indicate the estimated mode at current time $t{\,\in\,}\Nbb$.
    The blue circles indicate the expected pattern which is first to occur; in this example, $(3,1,4)$ has the highest first occurrence probability among any pattern in the collection $\Psi[t]$.
    Control input sequences are then scheduled according to Proposition~\ref{prop:mode_mpc_schedule}, shown in the green box.}
    \label{fig:modulation_mpc}
    \vspace{-10pt}
\end{figure}

\authorrevised{
Our controller architecture extends traditional uncertain system controllers (which borrow techniques from system identification and predictive control) via the incorporation of PLP.
We now provide in-depth discussions around each component in the following Sec.~\ref{sec:mode_id},~\ref{sec:pattern_learning}, and~\ref{sec:mode_control}, especially the concrete algorithms we choose to implement each component for the topology-switching network application to be demonstrated in~\sec{case_study}.
We again emphasize that our choices were made specifically for the MJS setup in~\sec{setup} and that other implementations of the controller architecture are possible.
For example, our paper~\cite{han23l4dc} describes a version for the problem of vehicle traffic congestion control, which explicitly includes a memory component to reduce the size of the table $\Ucal$.
}
\vspace{-10pt}


\authorrevised{
\section{Mode Process Identification}\label{sec:mode_id}
The Mode Process ID component estimates the current mode $\hat{\varphi}_{N[t]}^{(t)}$ and the TPM $\hat{P}^{(t)}$.
First, $\hat{\varphi}_{N[t]}^{(t)}$ is estimated using the \textit{consistent set narrowing} approach, which is a variation of nested convex body chasing used for model approximation in~\cite{ho21a}.
Second, $\hat{P}^{(t)}$ is estimated using empirical counts based on $\hat{\varphi}_{N[t]}^{(t)}$ and on estimates of the previous modes $\{\hat{\varphi}_{N[s]}^{(s)}\}_{s=0}^{t-1}$.

\subsection{Consistent Set Narrowing}
Because the distribution of the external noise process $\wvect[t]$ is unknown other than its norm bound, we employ consistent set narrowing, which checks the set of modes that are `consistent' with the state/control trajectories.
This method was employed in $(4)$ of~\cite{han20l4dc} and is similar to the more general nested convex body chasing approach described in~\cite{ho21a}, which was used for model approximation and selection for designing robust controls.

Denote the current mode-index as $n{\,\triangleq\,} N[t]{\,\in\,}\Nbb$.
By~\assum{timescale}, there are at most $\Delta T-1$ state and control values, $\xvect[T_{n}{\,:\,}t]$ and $\uvect[T_{n}{\,:\,}t]$, associated with a single mode $\varphi_{n}$.

\begin{definition}[Consistent Sets]\label{def:consistent_set}
    Over time, we construct a sequence of \textit{consistent sets} $\{\Ccal[t]\}_{t\in\Nbb}$ in the following way.
    For each $n{\,\in\,}\Nbb$, we initially set $\Ccal[T_n] {\,\triangleq\,}\Xcal$ because no observations about the current mode $\varphi_n$ have been made yet.
    Then for each $t{\,\in\,}(T_n, T_{n+1})$, if $\Ccal[t-1]{\,\neq\,}\varnothing$, a new consistent set is formed by retaining all modes $m{\,\in\,}\Ccal[t-1]$ from the previous consistent set $\Ccal[t-1]$ if each one-step value of state and control $(\xvect[r], \xvect[r+1], \uvect[t])$ satisfies the norm-boundedness condition of the noise $\wvect[t]$:
    \begin{align}\label{eq:consistent_set}
        &\Ccal[t] =\notag\\
        &\bigg\{m\in\Ccal[t-1]\,|\!\! \bigwedge\limits_{r=T_{n}}^{t-1}\!\!\mathds{1}\{\norm{\xvect[r\!+\!1] \!-\! A(m)\xvect[r] \!-\! B\uvect[r]}_{\infty} \!\leq\! \overline{w}\}\bigg\}
    \end{align}
\end{definition}

As~\eqn{consistent_set} is constructed for each $t{\,\in\,}\Nbb$, we update the mode estimate $\hat{\varphi}_{N[t]}^{(t)}\in\text{argmax}_{\zeta\in\Xcal}\hat{P}^{(t-1)}[\hat{\varphi}_{N[t-1]}^{(t-1)},\zeta]$ if $\abs{\Ccal[t]} > 1$; otherwise, we update $\hat{\varphi}_{N[t]}^{(t)} {\,\in\,}\Ccal[t]$.

\begin{figure}
    \centering
    \includegraphics[width=0.92\columnwidth]{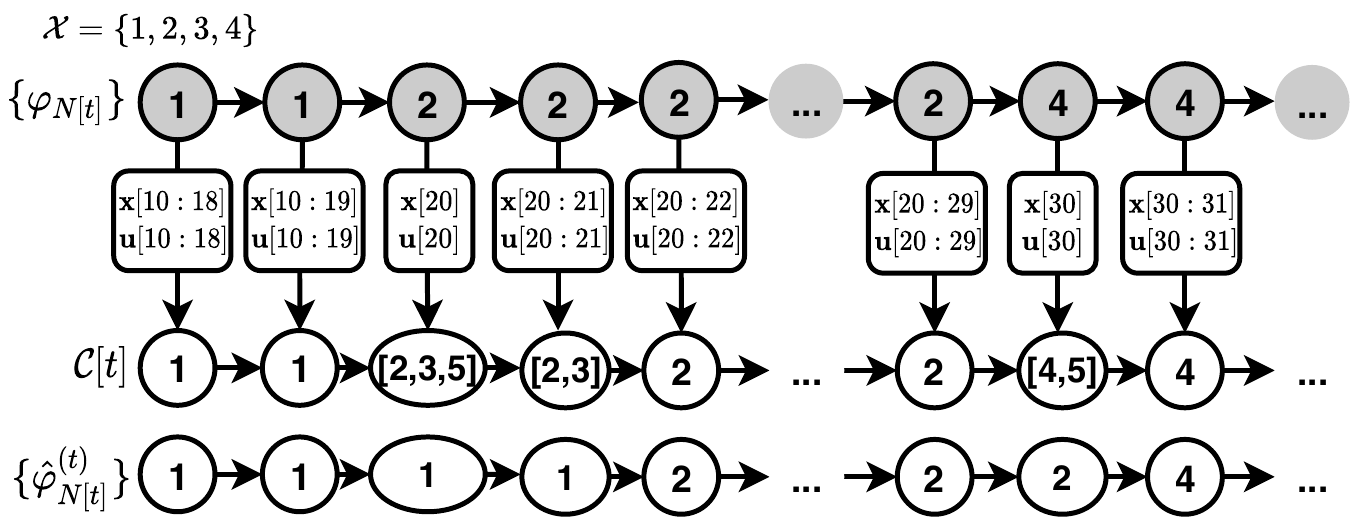}
    \caption{
    A visual diagram depicting Mode Process Identification.
    Here, $\Delta T{\,=\,}10$ and $M{\,=\,}5$.
    With $n\triangleq N[t]$, the upper row of gray circles denotes a realization $\{\varphi_n\}$ of the original unobservable mode process $\{\xi_n\}$.
    The middle row denotes the evolution of the consistent set, updated via~\eqn{consistent_set} using the state and control trajectories $\xvect[T_{n}{\,:\,}t]$, $\uvect[T_{n}{\,:\,}t]$.
    The estimate $\{\hat{\varphi}_{n}^{(t)}\}$ of the mode process is at the bottom row of white circles. 
    }
    \label{fig:mode_id_process}
    \vspace{-10pt}
\end{figure}

\begin{remark}
    One property of consistent set narrowing, also observed in nested convex body chasing approaches~\citep{ho21a}, is that at each time $t{\,\in\,}\Zbb^{\geq 0}$, the consistent set $\Ccal[t]$ always contains the true mode $\varphi_{N[t]}$.
    This is by definition of the consistent set, and the deterministic nature of the condition~\eqn{consistent_set} which defines the narrowing process (equivalent to verifying a simple linear inequality). 
    In the MJS literature, there have been notions of consistency similar to~\eqn{consistent_set} according to which unknown modes of MJS are estimated.
    For example,~\cite{schuurmans21} verifies consistency under the assumption that imperfect measurements $\yvect[t]{\,\neq\,}\xvect[t]$ of the state $\xvect[t]$ are collected.
    Thus, instead of using the state history $\xvect[0{\,:\,}t]$ directly, the consistency condition is designed around the collected measurements $\yvect[0{\,:\,}t]$ and propagated estimates of $\xvect[0{\,:\,}t]$ based on the initial condition $\xvect_0$ and the measurement equation.
    
    \textit{Mode detectability} is also a concept that has been studied;
    for instance, in~\cite{costa15}, the mode variable (analogous to $\xi_t{\,\in\,}\Xcal$ in our notation) emits its own signal (analogous to $\hat{\varphi}_t^{(t)}$ in our notation) independently of the system dynamics and the previous modes.
    In the consistent set narrowing approach, we obtain $\hat{\varphi}_n^{(t)}$ from the state and control trajectories $(\xvect[t],\xvect[t+1],\uvect[t])$; this estimate also changes with time as we collect more data about the trajectories.
\end{remark}

\subsection{Empirical Estimation of the TPM}
For any $n{\,\in\,}\Nbb$, the estimate of $\varphi_n$ is most accurate when the maximum possible amount of data from the system has been obtained to create the estimate, i.e., among all $t{\,\in\,}[T_n, T_{n+1})$, the value of $\hat{\varphi}_{n}^{(t)}$ is most accurate at time $t{\,=\,}T_{n+1}-1$.
For general $T_{N[t]}{\,<\,}t{\,<\,}T_{N[t]+1}$, $\hat{P}^{(t)}$ is estimated based on $\hat{\varphi}_{N[t]}^{(t)}$ 
and only the most accurate estimates of the previous modes $\{\hat{\varphi}_{N[s]}^{(T_{N[s]}-1)}\}_{s=0}^{t-1}$.
Thus, in the TPM estimation procedure, there is only one estimate associated each true mode $\varphi_n$. 
For simplicity of notation in this section only, we fix $n{\,\triangleq\,}N[t]$ and denote shorthand $\hat{\varphi}_{n'}{\,\equiv\,}\hat{\varphi}_{n'}^{(T_{n'}-1)}$ for $n'{\,<\,}n$ and $\hat{\varphi}_{n}{\,\equiv\,}\hat{\varphi}_{n}^{(t)}$.

If $t{\,=\,}T_n$ for some $n{\,\in\,}\Nbb$, estimating $\hat{P}^{(t)}$ given $\{\hat{\varphi}_{n'}\}_{n'=1}^{n}$ is straightforward.
By~\assum{knowns}, it is known which entries of the TPM are nonzero.
Thus, we initialize $\hat{P}^{(t)}$ to be an $M\times M$ matrix with a $1$ in the nonzero entries; when normalized, this corresponds to a stochastic matrix which has uniform distribution over the feasible transitions (e.g., $1/3$ probability each for a row with three nonzero entries) but for estimation purposes, we keep the estimate of the TPM unnormalized until the end of the simulation duration.
For each consecutive pair of transitions $(\hat{\varphi}_{n'},\hat{\varphi}_{n'+1})$ for $n'{\,\in\,}\{0,\cdots,n-1\}$, we take $\hat{P}^{(t)}[\hat{\varphi}_{n'},\hat{\varphi}_{n'+1}]{\,=\,}\hat{P}^{(t)}[\hat{\varphi}_{n'},\hat{\varphi}_{n'+1}]+ 1$.

If $T_n{\,<\,}t{\,<\,}T_{n+1}$ for some $n{\,\in\,}\Nbb$, we have two separate subcases.
If $\hat{\varphi}_{n}^{(t-1)}{\,=\,}\hat{\varphi}_{n}^{(t)}$, then we simply follow the approach above and compute $\hat{P}^{(t)}$ using the sequence $\{\hat{\varphi}_{n'}\}_{n'=1}^{n}$.
Otherwise, if $\hat{\varphi}_{n}^{(t-1)}{\,\neq\,}\hat{\varphi}_{N[t]}^{(t)}$, then we again follow the approach above and compute $\hat{P}^{(t)}$, but using the sequence $\{\hat{\varphi}_{n'}\}_{n'=1}^{n-1}$ instead.
To incorporate the mode estimate at current mode-index $n$, we first need to reset the TPM estimate of the last transition via $\hat{P}^{(t)}[\hat{\varphi}_{n-1},\hat{\varphi}_{n}^{(t-1)}]{\,=\,}\hat{P}^{(t)}[\hat{\varphi}_{n-1},\hat{\varphi}_{n}^{(t-1)}]{\,-\,}1$; then we update as usual $\hat{P}^{(t)}[\hat{\varphi}_{n-1},\hat{\varphi}_{n}^{(t)}]{\,=\,}\hat{P}^{(t)}[\hat{\varphi}_{n-1},\hat{\varphi}_{n}^{(t)}]{\,+\,}1$.
Once the mode sequence estimates have been processed until current time $t$, we update $\hat{P}^{(t)}$ such that each row is normalized to sum to $1$.

\begin{remark}
    The need for including Mode Process ID in the controller architecture~\fig{two_part_hierarchical_flow} is closely related to the notion of \textit{mode observability}, which has been studied extensively in the literature~\citep{vidal02,alessandri05,baglietto07,schuurmans21}.
    One common setup is that the measurements come from a (linear) noisy measurement equation such that $\yvect[t]{\,\neq\,}\xvect[t]$, and derive mode observability conditions from the imperfect observations $\yvect[t]$ of the state $\xvect[t]$.
    Also, the mode process is assumed to operate on the same timescale as the system dynamics.
    Compared to these methods, the algorithms we chose for implementing Mode Process ID hinge upon assumptions that simplify the mode observability problem.
    For example, in~\assum{knowns}, the state $\xvect[t]$ is observable and in~\assum{timescale}, we fix the mode switching times to be constant and deterministic rather than stochastic.
    
    We again emphasize that this is because the focus of our paper is on the impact of PLP on control design rather than mode observability, and we aimed to set up a simple scenario to show that our approach can be used when the system has uncertainties.
    Thus, not all of our assumptions are limiting; for example, compared to our approach,~\cite{vidal02} explicitly imposes that the external noise processes $\{\wvect[t]\}_t, \{\vvect[t]\}_t$ are Gaussian white and neither~\cite{vidal02} nor~\cite{alessandri05} consider the impact of control.
\end{remark}

\begin{remark}\label{rmk:simult_stab}
    We qualitatively discuss some conditions for mode observability in our specific implementation of Mode Process ID.
    First, the modes $\{A(1),\cdots, A(M)\}$ cannot be too ``similar'' to each other with respect to a certain metric $d$, (e.g., if $d(A(m_1),A(m_2)){\,<\,}\epsilon$ for some threshold $\epsilon{\,>\,}0$ and two distinct modes $m_1{\,\neq\,}m_2$ and $m_1,m_2{\,\in\,}\Xcal$).
    Second, when $\Delta T$ is too short, the consistent set may not converge to a single mode even if $d(A(m_1),A(m_2)){\,\geq\,}\epsilon$ for all pairs $(m_1,m_2){\,\in\,}\Xcal$ such that $m_1{\,\neq\,}m_2$.
    Rigorous derivation of these conditions for our specific use case are deferred to future work.
    This includes designing $d$ and $\epsilon$ for the consistent set narrowing approach, and deriving conditions on $\Delta T$ and the set $\{A(1), \cdots, A(M)\}$ for guaranteed convergence towards a singleton consistent set.
    Although these conditions are contingent upon our simplifying assumptions, they are expected to be similar to those derived in the aforementioned literature.
    For the purposes of our simulations in Sec.~\ref{sec:case_study}, $\Delta T$ and the different modes are empirically selected.
\end{remark}
\vspace{-10pt}
}

\authorrevised{
\section{Pattern-Learning for Prediction}\label{sec:pattern_learning}
}
\authorrevised{The Pattern-Learning component is implemented by using martingale theory to derive closed-form expressions about the pattern-occurrence quantities from~\prob{pattern_occurrence}, which are two important statistics that will aid with prediction on the mode process.
With martingales}, the resulting formulas yield better mathematical interpretation.
In scan statistics, martingales also allow for a more accurate test of experiment results than hypothesis testing~\citep{guerriero09}.

\subsection{Construction Based on Game Interpretation}\label{subsec:game_construction}

\begin{remark}\label{rmk:no_hat}
    We simplify the notation and remove the hats and the superscripts of $(t)$ in the estimated quantities throughout~\sec{pattern_learning} only.
    That is, for each $n$ and $t$ satisfying $N[t]{\,=\,}n$, we denote $\varphi_{n}{\,\equiv\,}\hat{\varphi}_{n}^{(t)}$, $P{\,\equiv\,}\hat{P}^{(t)}$, $\tau_{k|n}{\,\equiv\,} \hat{\tau}_{k|n}^{(t)}$, $\tau_{n}{\,\equiv\,} \hat{\tau}_{n}^{(t)}$, and $q_{k}{\,\equiv\,} \hat{q}_{k}^{(t)}$.
    Furthermore, we also remove the bracket $[t]$ in the pattern collection $\Psi[t]$ (see~\defin{pattern_collection_tv}), and use the notation $\Psi$ instead.
    However, we emphasize the understanding that the computation done at time $t$ uses the original estimates and the time-varying pattern collection.
\end{remark}

Note that there are constraints on the degrees of freedom on possible Markov chain sample path trajectories.
Thus, we take inspiration from~\cite{pozdnyakov08} and consider the occurrence of feasible augmented patterns up to two extra modes.

\begin{definition}[Augmented Pattern]\label{def:augmented_patterns}
    Suppose we are given a collection of patterns $\Psi$ (from~\defin{pattern_collection_tv}).
    An \textit{augmented pattern} $\boldsymbol{\gamma}$ corresponding to a pattern $\boldsymbol{\psi}_k{\,\in\,}\Psi$ is defined by prefixing two modes $m_1, m_2{\,\in\,}\Xcal$ such that the resulting sequence is feasible in the sense of~\defin{feasible}.
    We define the \textit{augmented collection}
    \begin{align}\label{eq:augmented_collection}
        \Gamma \triangleq \{\text{feasible } (m_1,m_2)\circ\boldsymbol{\psi}_k\hskip.1cm|\hskip.1cm m_1,m_2\in\Xcal; \boldsymbol{\psi}_k\in\Psi\}
    \end{align}
    to be the collection of augmented patterns, and we define $K_L{\,\in\,}\Nbb$ to be its cardinality.
    We enumerate each augmented pattern $\boldsymbol{\gamma}_{\ell}$ in the augmented collection $\Gamma$ using subscript $\ell{\,\in\,}\{1,\cdots, K_L\}$.
    Here, $\circ$ denotes the concatenation operation and each augmented pattern has length $L+2$.
\end{definition}

It is easier to solve for~\prob{pattern_occurrence} by conditioning on observing specific types of ending strings, formally defined below. 

\begin{definition}[Ending Strings]\label{def:ending_strings}
    Given the collection of patterns $\Psi$ and current mode-index $n{\,\in\,}\Nbb$, suppose we let the mode sequence $\{\xi_n, \xi_{n+1}, \cdots\}$ run until one of the patterns from $\Psi$ has been observed.
    Then an \textit{ending string} associated with pattern $\boldsymbol{\psi}_k{\,\in\,}\Psi$ terminates the mode process at mode-index $\tau_{n}{\,>\,}n$ if $\xi_{\tau_{n}-L+1:\tau_{n}}{\,=\,}\boldsymbol{\psi}_k$.
    We characterize two primary types of ending strings:
    \begin{itemize}[leftmargin=*]
        \setlength\itemsep{0em}
        \item \authorrevised{
        An \textit{initial-ending string} $\boldsymbol{\beta}$ occurs when part of an augmented pattern is observed immediately after the current mode.
        We classify initial-ending strings into two further subcases:
        \vspace{-5pt}
        \begin{itemize}[leftmargin=*]
            \item A \textit{Case $0$ initial-ending string} $\boldsymbol{\beta}{\,\triangleq\,}\boldsymbol{\psi}_k$ occurs when $\xi_{n+1:n+L}{\,=\,}\boldsymbol{\psi}_k$.
            Define $\Scal_I^{(0)}$ to be the set of Case 0 initial-ending strings with cardinality $ K_I^{(0)}{\,\in\,}\Nbb$.
            
            \item Let $m_1{\,\in\,}\Xcal$ be such that the above ending strings are feasible.
            A \textit{Case $1$ initial-ending string} $\boldsymbol{\beta}\triangleq (m_1)\circ \boldsymbol{\psi}_k$ occurs when $\xi_{n+1:n+L+1}=(m_1)\circ \boldsymbol{\psi}_k$.
            Define $\Scal_I^{(1)}$ to be the set of Case 1 initial-ending strings with cardinality $ K_I^{(1)}\in\Nbb$.
            \vspace{-5pt}
        \end{itemize}
        }

        \item \authorrevised{Let $m_1,m_2{\,\in\,}\Xcal$ be such that the above ending strings are feasible, and let $*$ be a placeholder for any feasible sequence of modes (see~\defin{feasible}) including the empty string.
        A \textit{later-ending string} $(*,m_1,m_2)\circ \boldsymbol{\psi}_k$ occurs when an augmented pattern is observed long after the current mode, i.e., when $\tau_{n} > n+L+1$ and $\xi_{\tau_{n}-L+1:\tau_{n}}{\,=\,} (m_1,m_2)\circ \boldsymbol{\psi}_k$.}
        Define $\Scal_L \triangleq\{(*)\circ\boldsymbol{\gamma}_{\ell}\,|\,\boldsymbol{\gamma}_{\ell}\in\Gamma\}$ to be the set of later-ending strings, with the same cardinality $K_L$ as $\Gamma$.
    \end{itemize}
    Define $\Scal_I{\,\triangleq\,}\Scal_I^{(0)}\cup\Scal_I^{(1)}$ with cardinality $K_I{\,=\,}K_I^{(0)}+K_I^{(1)}$, and let the \textit{set of ending strings} be $\Scal = \Scal_I\cup\Scal_L$.
    We enumerate each ending string $\boldsymbol{\beta}_s$ in $\Scal$ using the subscript $s{\,\in\,}\{1,\cdots,K_I+K_L\}$.
\end{definition}

\begin{figure}
    \centering
    \includegraphics[width=0.92\columnwidth]{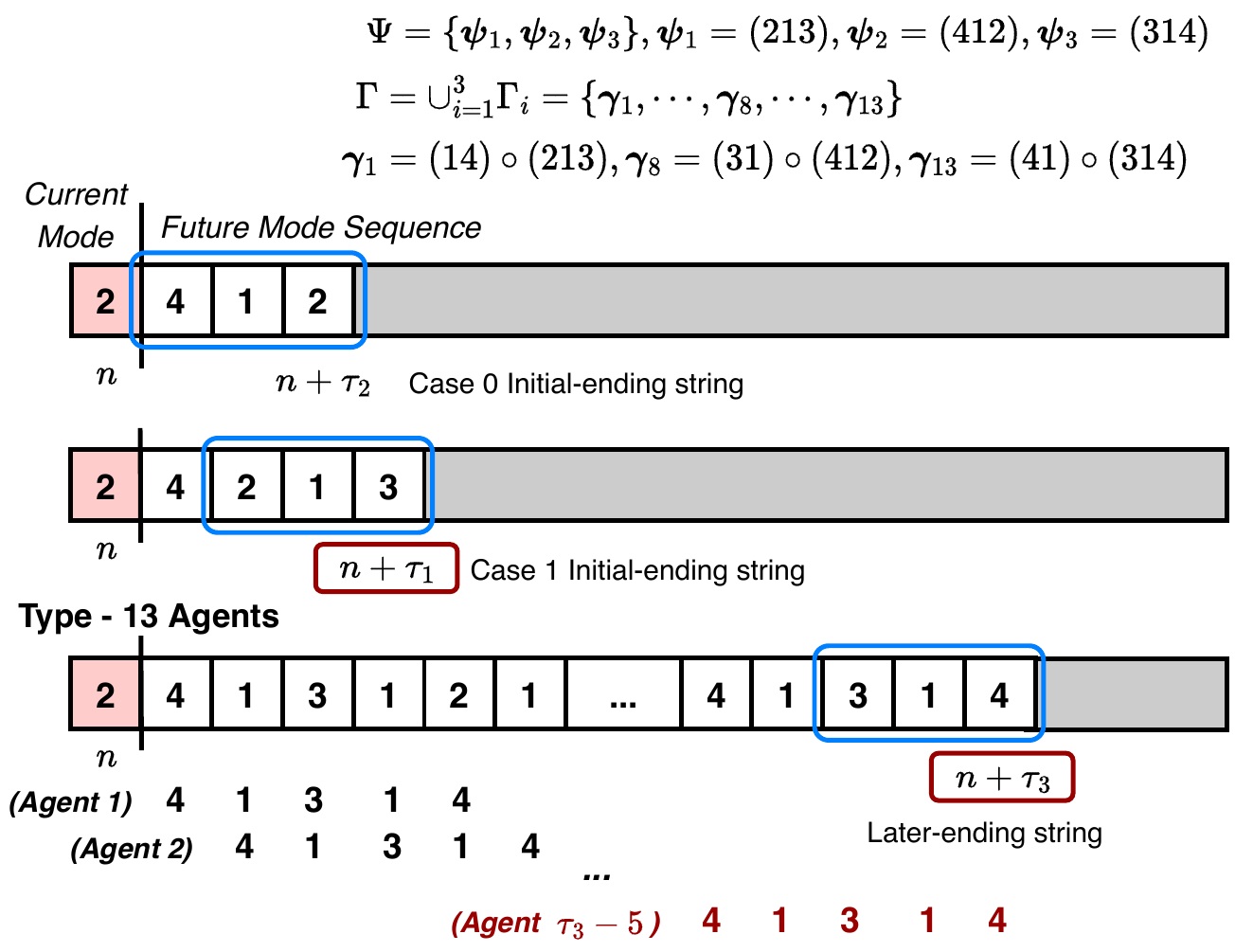}
    \caption{
    A visualization of the ending strings and agent-reward construction using the setup of~\exa{ending_string_example}.
    The red box marks the current mode-index $n\in\Nbb$, and each of the three sequences demonstrate the three different types of ending strings which terminate the mode process in the sense of~\defin{ending_strings}.
    The grey rectangles hide future modes which have not occurred because of termination.
    For the last case where $\boldsymbol{\gamma}_{13}$ terminates the mode process as a later-ending string, type-$13$ agents at mode-indices $1, 2, \cdots, \tau_3-5$ are shown.
    By the reward construction of~\defin{agents_rewards}, type-$13$ agent $\tau_3-5$ is the only agent who receives a nonzero reward.
    }
    \label{fig:martingale_proof}
\end{figure}

\vspace{-10pt}
\begin{example}[Ending Strings Construction]\label{ex:ending_string_example}
    We provide intuition behind the notation described by~\defin{ending_strings}.
    Let $M{\,=\,}4$, i.e., $\Xcal{\,=\,}\{1,2,3,4\}$, and let the (estimated) TPM $P$ be such that $P[m_1, m_2]{\,>\,}0$ for all $m_1,m_2{\,\in\,}\Xcal$ except when $m_1{\,=\,}m_2$ and when $(m_1,m_2){\,\in\,}\{(3,2), (2,3), (3,4), (4,3)\}$.
    The pattern collection consists of $K{\,=\,}3$ patterns $\Psi{\,=\,}\{\boldsymbol{\psi}_1, \boldsymbol{\psi}_2, \boldsymbol{\psi}_3\}$ of length $L{\,=\,}3$, with $\boldsymbol{\psi}_1{\,=\,}(213)$, $\boldsymbol{\psi}_2{\,=\,}(412)$, and $\boldsymbol{\psi}_3{\,=\,}(314)$.
    The augmented pattern collection is defined as $\Gamma{\,\triangleq\,}\cup_{i=1}^3\Gamma_i$ with
    $\Gamma_1{\,=\,}\{\boldsymbol{\alpha}\circ\authorrevised{\boldsymbol{\psi}_1}| \boldsymbol{\alpha}{\,\in\,}\{(14),(21),(24),(31),(41)\}\}$,
    $\Gamma_2{\,=\,}\{\boldsymbol{\alpha}\circ\authorrevised{\boldsymbol{\psi}_2}| \boldsymbol{\alpha}\in\{(12),(21),(31),(41),(42)\}\}$,
    $\Gamma_3{\,=\,}\{\boldsymbol{\alpha}\circ\authorrevised{\boldsymbol{\psi}_3}| \boldsymbol{\alpha}\in\{(21),(31),(41)\}\}$.
    The number of later-ending strings is $K_L{\,=\,}13$.
    Suppose the (estimated) current mode is $\varphi_n{\,=\,}2$.
    The set of feasible augmented Case 0 initial-ending strings is $\Scal_I^{(0)}{\,=\,}\{\boldsymbol{\psi}_2\}$ since $P[2,4]{\,>\,}0$.
    For Case 1 initial-ending strings, $\Scal_I^{(1)}{\,=\,}\{(1)\circ\boldsymbol{\psi}_1, (4)\circ\boldsymbol{\psi}_1, (1)\circ\boldsymbol{\psi}_2, (1)\circ\boldsymbol{\psi}_3, (2)\circ\boldsymbol{\psi}_2\}$.
    Thus, $K_I^{(0)}{\,=\,}1$ and $K_I^{(1)} = 4$.
\end{example}

\begin{definition}[Agents]\label{def:agents_rewards}
    Let $\Gamma$ be the augmented pattern collection associated with original collection $\Psi$ (see~\defin{augmented_patterns}).
    We introduce the notion of an \textit{agent}, which observes the mode process $\{\xi_n\}$ and accumulates \textit{rewards} at each mode-index with the goal of observing a pattern from $\Gamma$ (vicariously observing a pattern from $\Psi$).
    We refer to a \textit{type-$\ell$ agent} to be an agent which accumulates rewards by specifically observing the occurrence of $\boldsymbol{\gamma}_{\ell}{\,\in\,}\Gamma$ in $\{\xi_n\}$.
    At each mode-index $n{\,\in\,}\Nbb$, $K_L$ new agents, one for each type $\ell$, $\ell{\,\in\,}\{1,\cdots, K_L\}$, are introduced to the mode process; we refer to a type-$\ell$ agent which is introduced at mode-index $n$ as \textit{type-$\ell$ agent $n$}.
    A type-$\ell$ agent $n$ observes (estimated) mode realizations in the future sequence $\{\xi_{n+1}, \xi_{n+2}, \cdots, \}$ and accumulates rewards at a rate which is inversely-proportional to the action it took, starting with some arbitrary \textit{initial reward} $c_{\ell}{\,\in\,}\Rbb$.
    If $\varphi_n{\,=\,}m_1$, type-$\ell$ agent $n$ aims to observe the event $\{\xi_{n+1:n+L+1}{\,=\,}(m_2)\circ \boldsymbol{\psi}_k\}$.
    Otherwise, if $\varphi_n{\,\neq\,}m_1$, type-$\ell$ agent $n$ aims to observe the event $\{\xi_{n+1:n+L}{\,=\,}\boldsymbol{\psi}_k\}$.
\end{definition}

\begin{remark}\label{rmk:tau_augmented}
    It becomes necessary to distinguish the occurrence time of a pattern $\boldsymbol{\psi}_k$ from that of an augmented pattern $\boldsymbol{\gamma}_\ell \triangleq (m_1,m_2)\circ\boldsymbol{\psi}_k$.
    We define $\tau_{\ell|n}^a$ and $\tau_{n}^a$ to be the versions of~\eqn{pattern_def} and~\eqn{min_tau} for $\gamma_{\ell}\in\Gamma$.
\end{remark}

\begin{remark}\label{rmk:shift_indices}
    Due to the stationarity of $\{\xi_n\}$, the distributions of $\tau_{k|n_1}{\,-\,}n_1$ and $\tau_{k|n_2}{\,-\,}n_2$ are equivalent for each $k{\,\in\,}\{1,\cdots, K\}$, and any mode-indices $n_1, n_2{\,\in\,}\Nbb$, such that $\varphi_{n_1}{\,=\,}\varphi_{n_2}$.
    Likewise, the distributions of $\tau_{n_1}-n_1$ and $\tau_{n_2}-n_2$ are equivalent.
    For notation simplicity in the following presentation, we remove the subscript $n\in\Nbb$ in all variables, and use the above stationarity property to shift mode-indices to $n{\,=\,}0$ in variables such that the current mode is given by $\varphi_0$ instead of $\varphi_n$.
    Furthermore, we apply the shorthand notation to Definitions~\ref{def:pattern_times} and~\ref{def:min_tau_probs} such that $\tau_k{\,\equiv\,}\tau_{k|0}$ and $\tau{\,\equiv\,}\tau_{0}$; the notation for the augmented patterns (\remk{tau_augmented}) follow similarly as $\tau_{\ell}^a{\,\equiv\,}\tau_{\ell|0}^a$ and $\tau^a{\,\equiv\,}\tau_{0}^a$.
\end{remark}

\begin{definition}[Ending String Probabilities]\label{def:ending_string_prob}
    Define $\Pbb(\boldsymbol{\beta}_s)$ to be the probability that an ending string $\boldsymbol{\beta}_s{\,\in\,}\Scal$ terminates the mode process $\{\xi_n\}$ in the sense of~\defin{ending_strings}.
    For initial-ending strings $\boldsymbol{\beta}_s{\,\in\,}\Scal_I$ which is explicitly denoted as $(\beta_1, \cdots, \beta_{b_s})$ with length $b_s{\,\in\,}\Nbb$, we get $\Pbb(\boldsymbol{\beta}_s){\,=\,}P[\varphi_0,\beta_1]\prod_{j=1}^{b_s-1}P[\beta_j, \beta_{j+1}]$.
    We demonstrate how to compute $\Pbb(\boldsymbol{\beta}_s)$ for later-ending strings $\boldsymbol{\beta}_s{\,\in\,}\Scal_L$ in the following Sec.~\ref{subsec:theorems}, as part of solving~\prob{pattern_occurrence}.
\end{definition}

\begin{definition}[Gain Matrix]\label{def:gains_matrix_init_reward}
    Let $\boldsymbol{\beta}_s{\,\in\,}\Scal$ be an ending string which is explicitly denoted as $\boldsymbol{\beta}_s{\,\triangleq\,}(\beta_1, \cdots, \beta_{b_s}){\,\in\,}\Scal$ with length $b_s{\,\in\,}\Nbb$.
    Further let augmented pattern $\boldsymbol{\gamma}_{\ell}{\,\in\,}\Gamma$ be associated with original pattern $\boldsymbol{\psi}_k{\,\in\,}\Psi$, i.e., $\boldsymbol{\gamma}_{\ell}{\,\triangleq\,}(m_1,m_2)\circ \boldsymbol{\psi}_k$ for some $m_1,m_2{\,\in\,}\Xcal$.
    Then the \textit{total gain $W_{s\ell}$} accumulated over all type-$\ell$ agents from observing (partial) occurrences of $\boldsymbol{\gamma}_{\ell}$ in $\boldsymbol{\beta}_s$, is given by $W_{s\ell}{\,\triangleq\,}\sum_{j=1}^{\min(b_s-1,L+1)}D_j^{(1)}(\boldsymbol{\beta}_s,\boldsymbol{\gamma}_{\ell}){\,+\,}\sum_{j=1}^{\min(b_s-1,L)}D_j^{(2)}(\boldsymbol{\beta}_s,\boldsymbol{\gamma}_{\ell})$ with $D_i^{(1)}$ and $D_i^{(2)}$ defined based on the reward strategy from~\defin{agents_rewards}.
    \authorrevised{
    First,
    \begin{align*}
        D_i^{(1)}(\boldsymbol{\beta}_s,\boldsymbol{\gamma}_{\ell}){\,\triangleq\,}\left(P[m_1,m_2]P[m_2,\psi_{k,1}]\prod_{j=2}^{i-1}P[\psi_{k,j-1}, \psi_{k,j}]\right)^{-1}
    \end{align*}
    if $\beta_{b_s-i}{\,=\,}m_1$ and $\beta_{b_s-i+1}{\,=\,}m_2, \beta_{b_s-i+j}{\,=\,}\psi_{k,j-1}$ for all $j{\,\in\,}\{2, \cdots, i\}$; else, $D_i^{(1)}(\boldsymbol{\beta}_s,\boldsymbol{\gamma}_{\ell}){\,=\,}0$.
    Second,
    \begin{align*}
        D_i^{(2)}(\boldsymbol{\beta}_s,\boldsymbol{\gamma}_{\ell}){\,\triangleq\,}\left(P[\beta_{b_s-i},\psi_1]\prod_{j=2}^{i}P[\psi_{k,j-1},\psi_j]\right)^{-1}
    \end{align*}
    if $\beta_{b_s-i}{\,\neq\,}m_2$ and $\beta_{b_s-i+j}{\,=\,}\psi_j$ for all $j{\,\in\,}\{1,\cdots,i\}$; else, $D_i^{(2)}(\boldsymbol{\beta}_s,\boldsymbol{\gamma}_{\ell})=0$.
    }
    A \textit{gain matrix} $W {\,\in\,}\Rbb^{(K_I+K_L)\times K_L}$ is constructed with entries $W_{s\ell}$ for each pair of $\boldsymbol{\beta}_s{\,\in\,}\Scal$ and $\boldsymbol{\gamma}_{\ell}{\,\in\,}\Gamma$.
\end{definition}

\begin{definition}[Cumulative Net Reward]\label{def:cum_net_rewards}
    The expected \textit{type-$\ell$ cumulative net reward} over all type-$\ell$ agents by mode-index $\tau$ is defined $\Ebb[R_{\tau}^{(\ell)}]{\,\triangleq\,}c_{\ell}([\Pbb(\boldsymbol{\beta}_1),\cdots,\Pbb(\boldsymbol{\beta}_{K_I+K_L})]W_{\cdot, \ell} - \Ebb[\tau])$, where the $\Pbb(\boldsymbol{\beta}_s)$ are the probabilities from~\defin{ending_string_prob} and $W_{\cdot, \ell}$ denotes the $\ell$th column of the gain matrix (see~\defin{gains_matrix_init_reward}). 
    Correspondingly, the expected \textit{cumulative net reward} over all agents by mode-index $\overline{n}$ is defined as $R_{\overline{n}} \triangleq \sum_{\ell=1}^{K_L} R_{\overline{n}}^{(\ell)}$, and
    \begin{align}\label{eq:cumu_reward_gain_matrix_relation}
        &\hskip.1cm\Ebb[R_{\tau}] \!=\![\Pbb(\boldsymbol{\beta}_1)\cdots\Pbb(\boldsymbol{\beta}_{K_I+K_L})]W\cvect \!-\!\left(\sum\limits_{\ell=1}^{K_L}c_{\ell}\right)\Ebb[\tau]
    \end{align}
    where $\cvect{\,\triangleq\,}[c_1, \cdots, c_{K_L}]^{\top}$ are the initial rewards (\defin{agents_rewards}).
\end{definition}
\vspace{-10pt}

\authorrevised{
\subsection{Solving the Pattern-Occurrence Problem}\label{subsec:theorems}
}
We are now ready to use our construction to present our main results, which address the questions in~\prob{pattern_occurrence}.

\begin{theorem}[Expected Time of Occurrence]\label{thm:expected_tau}
    Denote $\tau$ as in~\remk{shift_indices} with (estimated) current mode $\varphi_{0}$ for the collection $\Psi$ from~\defin{patterns} and corresponding augmented collection $\Gamma$.
    Then
    \begin{align}\label{eq:compute_tau_mc}
        \Ebb[\tau] = \frac{1}{\sum\limits_{\ell=1}^{K_L}c_{\ell}^{*}}\Bigg[\Bigg(1 - \sum\limits_{s=1}^{K_I} \Pbb(\boldsymbol{\beta}_s)\Bigg)+ \sum\limits_{s=1}^{K_I} \Pbb(\boldsymbol{\beta}_s)\sum\limits_{\ell=1}^{K_L}W_{s\ell}c_{\ell}^{*}\Bigg]
    \end{align}
    where $\boldsymbol{\gamma_{\ell}}{\,\in\,}\Gamma$, $\boldsymbol{\beta}_s{\,\in\,}\Scal$, $\Pbb(\boldsymbol{\beta}_s)$ is from~\defin{ending_string_prob}, $W$ is from~\defin{gains_matrix_init_reward}, and $\cvect^{*}{\,\in\,}\Rbb^{K_L}$ is the vector of initial rewards (see~\defin{gains_matrix_init_reward}) such that $\sum_{\ell=1}^{K_L}W_{s\ell}c_{\ell}^{*}{\,=\,}1$ for all $s{\,\in\,}\{K_I{\,+\,}1, \cdots, K_I+K_L\}$.
\end{theorem}
\begin{proof}
    Because the Markov chain is irreducible and finite-state, $\Ebb[\tau^a_{\ell}]{\,<\,}\infty$, for each $\tau_{\ell}^a$ defined in~\remk{tau_augmented}.
    Note that $\tau_k{\,=\,}\min_{\boldsymbol{\gamma}_{\ell}\in\Gamma_k} \tau^{a}_{\ell}$, where $\Gamma_k$ is the subset of $\Gamma$ containing augmented patterns $\boldsymbol{\gamma}{\,\triangleq\,}(m_1,m_2)\circ\boldsymbol{\psi}_k$ corresponding to original pattern $\boldsymbol{\psi}_k{\,\in\,}\Psi$.
    We have that $\tau{\,\triangleq\,} \min_{k}\tau_{k}$, and by~\defin{min_tau_probs}, we also have $\Ebb[\tau]{\,<\,}\infty$.
    %
    By the construction of the gain matrix $W$ and the fact that linear combinations of martingales are martingales, both $\{R_{n\wedge\tau^{a}_{\ell}}^{(\ell)}\}_{n\in\Nbb}$ and $\{R_{n\wedge\tau}\}_{n\in\Nbb}$ are martingales.
    This implies that $\Ebb[R_{\tau^{a}_{\ell}}^{(\ell)}]{\,<\,}\infty$ since $\Ebb[\tau^{a}_{\ell}]{\,<\,}\infty$. 
    Furthermore, $\Ebb[R_{\tau}]{\,<\,}\infty$ because $\tau{\,\leq\,}\tau^{a}_{\ell}$ for all $\ell$.
    Define the set $\Omega_n^{(\ell)}{\,\triangleq\,}\{\omega{\,\in\,}\Omega | n{\,<\,}\tau^{a}_{\ell}\}$.
    By Doob's martingale convergence theorem and the triangle inequality, $\lim_{n\to\infty}\int_{\Omega_n^{(\ell)}} \abs{R_n^{(\ell)}(\omega)} d\Pbb(\omega) = 0$, which implies $\{R_{n\wedge\tau^{a}_{\ell}}^{(\ell)}\}$ is uniformly-integrable over $\Omega_n^{(\ell)}$.
    Thus, $\{R_{n\wedge\tau}\}$ is uniformly-integrable over $\Omega_n{\,\triangleq\,} \{\omega{\,\in\,}\Omega{\,|\,}n{\,<\,}\tau\} {\,\subseteq\,} \cap_{\ell=1}^{K_L} \Omega_n^{(\ell)}$.
    With the above conditions satisfied, we apply the optional stopping theorem to the stopped process $\{R_{n\wedge\tau}\}$, which implies $\Ebb[R_{\tau}]{\,=\,}\Ebb[R_0]$.
    Note $\Ebb[R_0]{\,=\,}0$ by the construction of~\defin{cum_net_rewards}.
    After choosing the initial rewards $\cvect^{*}$ as in the theorem statement, and substituting into~\eqn{cumu_reward_gain_matrix_relation}:
    \begin{align*}
        \Ebb[R_{\tau}]\!&=\!\sum\limits_{s=1}^{K_I}\Pbb(\boldsymbol{\beta}_s)\sum\limits_{\ell=1}^{K_L}W_{s\ell}c_{\ell}^{*}\!+\!\left(1\!-\!\sum\limits_{s=1}^{K_I}\Pbb(\boldsymbol{\beta}_s)\right)\!-\!\sum\limits_{\ell=1}^{K_L} c_{\ell}\Ebb[\tau]
    \end{align*}
    Rearranging the terms to isolate $\Ebb[\tau]$ yields~\eqn{compute_tau_mc}.
\end{proof}

This addresses the first question in~\prob{pattern_occurrence}.
To address the second question, we use the following theorem, which also addresses the computation of $\Pbb(\boldsymbol{\beta}_s)$ for any later-ending string $\boldsymbol{\beta}_s{\,\in\,}\Scal_L$ (see~\defin{ending_string_prob}).

\begin{theorem}[First-Occurrence Probabilities]\label{thm:first_occurrence}
    In addition to the setup of~\thm{expected_tau}, explicitly denote as ending string  as $\boldsymbol{\beta}_s{\,\triangleq\,}(\beta_1, \cdots, \beta_{b_s}){\,\in\,}\Scal$ to have length $b_s{\,\in\,}\Nbb$. 
    Then the (estimated) first-occurrence probabilities $\{q_k\}$ (see~\defin{min_tau_probs} and~\remk{no_hat}) are given by $q_k{\,=\,}\sum_{\boldsymbol{\beta}_s\in\Scal} \Pbb(\boldsymbol{\beta}_s)\mathds{1}\{\beta_{b_s-L+1:b_s}{\,=\,}\boldsymbol{\psi}_k\}$.
\end{theorem}
\begin{proof}
    Rearranging the terms of~\eqn{cumu_reward_gain_matrix_relation} yields:
    \begin{align}\label{eq:first_prob_system_eqns}
        \sum\limits_{s=1}^{K_I}\Pbb(\boldsymbol{\beta}_s)\sum\limits_{\ell=1}^{K_L}W_{s\ell}c_{\ell} = -\sum\limits_{s=K_I+1}^{K_I+K_L}\Pbb(\boldsymbol{\beta}_s)\sum\limits_{\ell=1}^{K_L}W_{s\ell}c_{\ell} +\sum\limits_{\ell=1}^{K_L} c_{\ell}\Ebb[\tau]
    \end{align}
    We are given $\Ebb[\tau]$ from~\thm{expected_tau}, and $\Pbb(\boldsymbol{\beta}_s)$ can be computed via~\defin{ending_string_prob} when $\boldsymbol{\beta}_s{\,\in\,}\Scal_I$.
    For $s{\,\in\,}\{K_I+1,\cdots,K_I+K_L\}$, choose one of $K_L$ vectors $\cvect\in\{\evect_1,\cdots,\evect_{K_L}\}$ (where $\evect_i$ is the $i$th standard basis vector of $\Rbb^{K_L}$) to substitute into~\eqn{first_prob_system_eqns} and construct $K_L$ different equations.
    Solve the resulting linear system for the $K_L$ unknowns $\{\Pbb(\boldsymbol{\beta}_{K_I+1}),\cdots,\Pbb(\boldsymbol{\beta}_{K_I+K_L})\}$.
    By~\defin{min_tau_probs}, $q_k{\,\triangleq\,}\Pbb(\tau{\,=\,}\tau_k){\,=\,}\sum_{\boldsymbol{\beta}_s\in\Scal} \Pbb(\boldsymbol{\beta}_s)\Pbb(\tau{\,=\,}\tau_k\,|\,\boldsymbol{\beta}_s)$, where we denote shorthand $\Pbb(\tau{\,=\,}\tau_k\,|\,\boldsymbol{\beta}_s)$ to be the probability of $\boldsymbol{\psi}_k$ being the first pattern observed at mode-index $\tau$ given $\boldsymbol{\beta}_s$ is the ending string which terminated the mode process in the sense of~\defin{ending_strings}.
    Clearly, $\Pbb(\boldsymbol{\psi}_k\,|\,\boldsymbol{\beta}_s){\,=\,}1$ if $\beta_{b_s-L+1:b_s}{\,=\,}\boldsymbol{\psi}_k$ holds, otherwise it is $0$.
    We thus obtain the desired equation.
\end{proof}

\begin{remark}
    In order to fit the closed-form expressions of Theorems~\ref{thm:expected_tau} and~\ref{thm:first_occurrence} into the original architecture described throughout~\sec{framework}, we unsimplify the notation from~\remk{no_hat} and~\remk{shift_indices} for general time $t{\,\in\,}\Nbb$ and corresponding mode-index $n{\,\triangleq\,}N[t]$.
    This yields the original time-dependent pattern-occurrence quantities desired in~\prob{pattern_occurrence}.
    Namely, with estimated current mode $\hat{\varphi}_{n}^{(t)}$ and TPM $\hat{P}^{(t)}$, the estimated expected minimum occurrence time $\Ebb[\hat{\tau}_{n}^{(t)}]$ is the $\Ebb[\tau]$ computed from~\thm{expected_tau}, while the estimated first occurrence probabilities $\{\hat{q}_k^{(t)}\}$ are the $\{q_k\}$ computed from~\thm{first_occurrence}.
\end{remark}

\vspace{-10pt}


\authorrevised{
\section{Control Law Design}\label{sec:mode_control}

In this section, we tie the pattern-occurrence quantities developed in~\sec{pattern_learning} into our choice of implementation for the Control Law Design component.
One well-known control method that explicitly incorporates predictions is \textit{model predictive control (MPC)}, and so we use principles similar to MPC to schedule control policies in advance (see Proposition~\ref{prop:mode_mpc_schedule}).
For the purposes of our dynamic topology network case study in Sec.~\ref{sec:case_study}, we also discuss non-predictive Control Law Design using the novel system level synthesis approach~\citep{wang18,anderson19}, including a topology-robust version~\citep{han20l4dc} and a data-driven version~\citep{xue21,carmen22}.

\subsection{Incorporating Predictions}\label{subsec:mpc}
We implement a table $\Ucal$ that maps patterns of interest to the optimal control sequences we designed for them in our experiment so far (see Proposition~\ref{prop:mode_mpc_memory}); this also includes explicit state and control trajectories.
This implementation was inspired by \textit{episodic memory}~\citep{lengyel07} which can be added to learning-based control methods (e.g., reinforcement learning) to recall specific experiences and their rewards~\citep{blundell16}.
Our table $\Ucal$ is implemented according to Proposition~\ref{prop:mode_mpc_memory} and its entries are updated in two ways: 1) the control law is updated in an entry for an existing pattern, or 2) a new entry is created for a newly-observed pattern $\boldsymbol{\psi}$ at time $t$, where $\boldsymbol{\psi}{\,\in\,}\Psi[t{\,+\,}1]$ but $\boldsymbol{\psi}{\,\not\in\,}\Psi[t]$.
We describe the control law synthesis and update procedures in the following Sec.~\ref{subsec:sls}.

For the prediction component, we specifically recall model predictive control (MPC).
Standard MPC for discrete-time linear dynamics seeks to predict a future sequence of controls $\{\uvect[t], \uvect[t+1], \cdots, \uvect[t+H]\}$ which minimizes some cost functional at each timestep $t{\,\in\,}\Nbb$, for some prediction horizon $H{\,\in\,}\Nbb$.
Once the first control input $\uvect[t]$ is applied to the system, the procedure is repeated at the next time $t+1$.
Although intuitive, incorporating both short-term and long-term predictions for online control have been proven to be beneficial, even when the system to be controlled is perturbed by either random and adversarial disturbances~\citep{chen15wierman}; in~\cite{yu20wierman}, this is demonstrated explicitly with the linear quadratic regulator.
For the concreteness of this paper, we are inspired by the methods of~\cite{park02} and~\cite{lu13}, which discuss MPC for MJS, and we extend their approaches to our setting from~\sec{setup}.

We remark that $H$, like prediction horizon $L$ for the mode process, is a user-chosen hyperparameter; one reasonable choice could be to make it time-varying and set it equal to $\Delta T - (t-T_{N[t]})$ at each $t$.
Given the estimated current mode $m{\,\triangleq\,}\hat{\varphi}_{N[t]}^{(t)}$, the cost function we seek to optimize is the following mode-dependent quadratic cost function:
\begin{align}\label{eq:quadratic_cost}
    J(t,m) &\triangleq \sum\limits_{s=t}^{H} (\xvect[s]^{\top}Q(m)\xvect[s] + \uvect[s]^{\top}R(m)\uvect[s])\notag\\
    &\hskip1cm + \xvect[H]^{\top}Q_f(m)\xvect[H]
\end{align}

The main distinction is that the prediction part of MPC is done on the estimated mode process instead of the system dynamics.
Let $t{\,\in\,}\Nbb$ and $n{\,\triangleq\,}N[t]$, and suppose the consistent set narrowing approach of Sec.~\ref{subsec:mode_id} estimates the current mode to be $\hat{\varphi}_{n}^{(t)}$.
Again, by~\assum{timescale}, there are at most $\Delta T-1$ state and control observations $\xvect[T_{n}{\,:\,}t]$ and $\uvect[T_{n}{\,:\,}t]$ associated with each mode $\varphi_{n}$.
Thus, for the control input $\uvect[t] = K(t,\hat{\varphi}_{n}^{(t)})\xvect[t]$ at time $t$, the gain $K(t,\hat{\varphi}_{n}^{(t)}){\,\in\,}\Rbb^{n_x\times n_u}$ associated specifically with mode $\hat{\varphi}_{n}^{(t)}$ can be designed using standard linear optimal control tools such as LQR minimization.

\subsection{System Level Synthesis}\label{subsec:sls}
For the purposes of this paper (especially for our case study in~\sec{case_study}), we employ the novel \textit{system level synthesis} (SLS)~\citep{wang18,anderson19} approach for distributed disturbance-rejection in linear discrete-time network systems with static topologies $\Gcal{\,\triangleq\,}(\Vcal, \Ecal)$, expressed as
\begin{align}\label{eq:linear_dt_static}
    \xvect[t+1] = A\xvect[t] + B\uvect[t] + \wvect[t]
\end{align}
%
The standard state-feedback control law for systems of this form is given by $\uvect[t]{\,=\,}K\xvect[t]$ and in $z$-transform expression, the resulting closed-loop system is given by $\xvect{\,=\,}(zI - A - BK)^{-1}\wvect$.
However, for large-scale systems (i.e., large-dimensional matrices $A$ and $B$), optimizing over the transfer function $(zI - A - BK)^{-1}$ by solving for $K$ is difficult.
Thus, a key feature of SLS is that it reparametrizes the control problem: instead of designing just the open-loop feedback gain $K$, SLS designs for the entire closed-loop system via response maps $\boldsymbol{\Phi}{\,\triangleq\,}\{\boldsymbol{\Phi}_x,\boldsymbol{\Phi}_u\}$ such that $\xvect[0{\,:\,}t]{\,=\,}\boldsymbol{\Phi}_x\wvect[0{\,:\,}t]$ and $\uvect[0{\,:\,}t]{\,=\,}\boldsymbol{\Phi}_u\wvect[0{\,:\,}t]$, where $\wvect[t]$ is an additive external disturbance.
\begin{lemma}
    For the linear, discrete-time static dynamics~\eqn{linear_dt_static}, the following are true.
    First, the affine subspace described by 
    \begin{align}\label{eq:achievable_sls}
        \begin{bmatrix}
            I - Z\hat{A} & -Z\hat{B}
        \end{bmatrix}\begin{bmatrix}
            \boldsymbol{\Phi}_x\\ \boldsymbol{\Phi}_u
        \end{bmatrix} = I
    \end{align}
    parametrizes all possible system responses $\boldsymbol{\Phi}$, where $\hat{A}\triangleq\text{blkdiag}(A,\cdots,A,\textbf{0}){\,\in\,}\Rbb^{Hn_x\times Hn_x}$, $\hat{B}$ is defined similarly, $Z$ is the block-downshift operator, $n_x{\,\in\,}\Nbb$ is the state dimension, and $H{\,\in\,}\Nbb$ is a chosen finite horizon over which control is performed.
    Second, for any $\boldsymbol{\Phi}$ which satisfies the condition in~\eqn{achievable_sls}, the feedback gain $K{\,\triangleq\,}\boldsymbol{\Phi}_u\boldsymbol{\Phi}_x^{-1}$ achieves the desired internally-stabilizing system response.
\end{lemma}

The state-feedback controller is then implemented with:
\begin{align}\label{eq:sf_implement}
    \hat{\xvect}[t] &= \sum\limits_{s=2}^H \Phi_x[s] \hat{\wvect}[t+1-s],\hskip.1cm \hat{\wvect}[t] = \xvect[t] - \hat{\xvect}[t]\notag\\
    \uvect[t] &= \sum\limits_{s=1}^H \Phi_u[s] \hat{\wvect}[t+1-s]
\end{align}
where $\hat{\wvect}$ is the controller's \textit{internal state} and $\hat{\xvect}$ is the controller's estimate of the state.
This form also makes SLS more suitable for distributed and localized control law design in large-scale linear systems, and so $\boldsymbol{\Phi}$ is often implemented as $\boldsymbol{\Phi}^{(i)}{\,\triangleq\,}\{\Phi_{x}^{(i)}[s], \Phi_{u}^{(i)}[s]\}$ for each node $i{\,\in\,}\Vcal$ and its local subsystem $\Lcal_{i,h}$.
Here, $s{\,\in\,}\{1,\cdots, H\}$ is the index of the \textit{spectral component}, $\Lcal_{i,h}$ is the set of all $j{\,\in\,}\Vcal$ which is within $h{\,\in\,}\Nbb$ edges away from $i$, and $h$ is some number of \textit{hops}.
Both time horizon $H$ and number of neighboring hops $h$ are parameters chosen by design based on properties such as the scale and topology of $\Gcal$.

We can also extend SLS to account for dynamic topologies $\Gcal(m){\,\triangleq\,}(\Vcal, \Ecal(m))$ for $m{\,\in\,}\Nbb$ representing the index of the topology; this was done in~\cite{han20l4dc}.
Let $\boldsymbol{\Phi}_m^{(i,t)}{\,\triangleq\,}\{\Phi_{x,m}^{(i,t)}[s], \Phi_{u,m}^{(i,t)}[s]\}$ define the $i$th local response map $\boldsymbol{\Phi}^{(i)}$ which is created specifically for topology $m{\,\in\,}\{1,\cdots, M\}$.
As we demonstrate for our case study in~\sec{case_study}, the mode in our original dynamics~\eqn{plant_dynamics} corresponds to the index of the current topology the system is in.
\textit{Topology-robust SLS} essentially attempts to design a single $\{\Phi_x, \Phi_u\}$ response that can simultaneously stabilize multiple topologies (i.e., distinct $A$ matrices). 
Conditions for \textit{simultaneous stabilization} for a collection of discrete-time LTI systems have been studied extensively in past literature: some results (e.g.,~\cite{blondel92}) express the condition by ensuring that the closed-loop transfer function between every possible plant-controller pair does not have any pole-zero cancellations, while others (e.g.,~\cite{cao99}) derive conditions based on the algebraic Riccati equation.
To keep our discussion focused, we do not state these conditions here (see~\remk{simult_stab}).

To be able to use PLP with the SLS approach, we require a formulation of SLS which is driven by data.
Towards that end, we leverage \textit{data-driven SLS}~\citep{xue21,carmen22}, which extends traditional SLS using a characterization based on Willems' fundamental lemma~\citep{willems97}, which parametrizes state
and input trajectories based on past trajectories under the conditions of persistence of excitation.
Define the Hankel matrix
\begin{align*}
    \hat{H}_r(\xvect[0:H]){\,\triangleq\,}\begin{bmatrix}
        \xvect[0] & \xvect[1] & \cdots & \xvect[H-r]\\
        \xvect[1] & \xvect[2] & \cdots & \xvect[H-r+1]\\
        \vdots & \vdots & \ddots & \vdots\\
        \xvect[r-1] & \xvect[r] & \cdots & \xvect[H-1]
    \end{bmatrix}
\end{align*}
for finite time horizon $H$ and some $r{\,\in\,}\Nbb$.
We say the finite-horizon state trajectory $\xvect[0:H]$ is \textit{persistently-exciting} of order $r$ if $\hat{H}_r(\xvect[0:H])$ is full rank.
In the data-driven formulation of SLS, the achievable subspace described by~\eqn{achievable_sls} can be equivalently written as the set
\begin{align}\label{eq:achievable_data_sls}
    \left\{\begin{bmatrix}
        \hat{H}_H(\xvect[0{\,:\,}H])\\ \hat{H}_H(\uvect[0{\,:\,}H])
    \end{bmatrix}G \hskip.1cm\bigg|\hskip.1cm G \text{ s.t. } \hat{H}_1(\xvect[0{\,:\,}H])G = I\right\}
\end{align}

Now, let $n{\,\in\,}\Nbb$ and $n'{\,\in\,}\Nbb$, $n'{\,>\,}n$, be such that at times $T_n$ and $T_{n'}$, the system~\eqn{plant_dynamics} have switched to the same mode $m{\,\in\,}\Xcal$.
For our PLP approach, the state/control trajectories $\{\xvect[T_{n-1}{\,:\,}T_n-1],\uvect[T_{n-1}{\,:\,}T_n-1]\}$ and $\{\xvect[T_{n'-1}{\,:\,}T_{n'}-1],\uvect[T_{n'-1}{\,:\,}T_{n'}-1]\}$ can be collectively used to design the optimal control law for mode $m$, i.e., we use horizon $T_{n-1}{\,:\,}T_n-1$ in place of $[0{\,:\,}H]$ in~\eqn{achievable_data_sls}.
To implement memory, we store (in $\Ucal$) previous trajectories of the system corresponding to the same mode, and continue to append to it as the simulation progresses. 
To apply Proposition~\ref{prop:mode_mpc_schedule}, SLS is run more than once to compute a new $\boldsymbol{\Phi}$ for every new estimated mode $m{\,\triangleq\,}\hat{\varphi}_{n}^{(t)}$, hence the dependence of $\boldsymbol{\Phi}_m^{(i,t)}$ on time $t{\,\in\,}\Nbb$.
By Proposition~\ref{prop:mode_mpc_memory}, the $\boldsymbol{\Phi}_m^{(i,t)}$ are stored and updated over time in the table $\Ucal$, then used to compute $\uvect[t]$ via~\eqn{sf_implement}.
\vspace{-10pt}
}


\section{Case Study: Topology-Switching Network}\label{sec:case_study}
Controlling networks that undergo parametric and/or topological changes (e.g., due to faults or connectivity changes of mobile agents) is an important and widely-studied problem in large-scale networked systems.
In the recent literature, an adaptive, consensus-based control scheme for complex networks with time-varying, switching network topology was discussed in~\cite{chung13}.
Distributed target-detection and tracking using a dynamic sensor network was studied in~\cite{bandyop17}, while~\cite{saboori15} described fault-tolerance against actuator failures in a multiagent system connected by a switching topology network.

For the purposes of this paper, we demonstrate the proposed controller architecture to the following extension of~\eqn{plant_dynamics}, which switches among a finite number of different topologies $\Gcal(m) \triangleq (\Vcal, \Ecal(m)), m\in\{1,\cdots,M\}, M{\,\in\,}\Nbb$.
\begin{align}\label{eq:power_plant_dynamics}
    \xvect_i[t+1] &= A_{ii}(\xi_{N[t]})\xvect_i[t]\notag\\
    &+ \sum\limits_{j\in\Ncal_i(\xi_{N[t]})}A_{ij}(\xi_{N[t]})\xvect_j[t] + B_i\uvect[t] + \wvect_i[t]
\end{align}
Here, $n_s{\,\triangleq\,}\abs{\Vcal}$, $i{\,\in\,}\{1,\cdots, n_s\}$, the neighboring nodes of subsystem $i$ are $\Ncal_i(m){\,\triangleq\,}\{j{\,\in\,}\Vcal{\,:\,}(i,j){\,\in\,}\Ecal(m)\}$, and $A(m) \triangleq [A_{ij}(m)]{\,\in\,}\Rbb^{n_x\times n_x}$ for each topology $m{\,\in\,}\{1,\cdots,M\}$.
The assumptions from Sec.~\ref{sec:setup} still hold, and the mode process $\{\xi_n\}$ is the index of the current topology at time $t{\,\in\,}\Nbb$ with $N[t]$ being the number of topology changes made by time $t$.

\subsection{Experiment Setup}\label{subsec:case_study_setup}
\begin{figure}
    \begin{center}
        \includegraphics[width=\columnwidth]{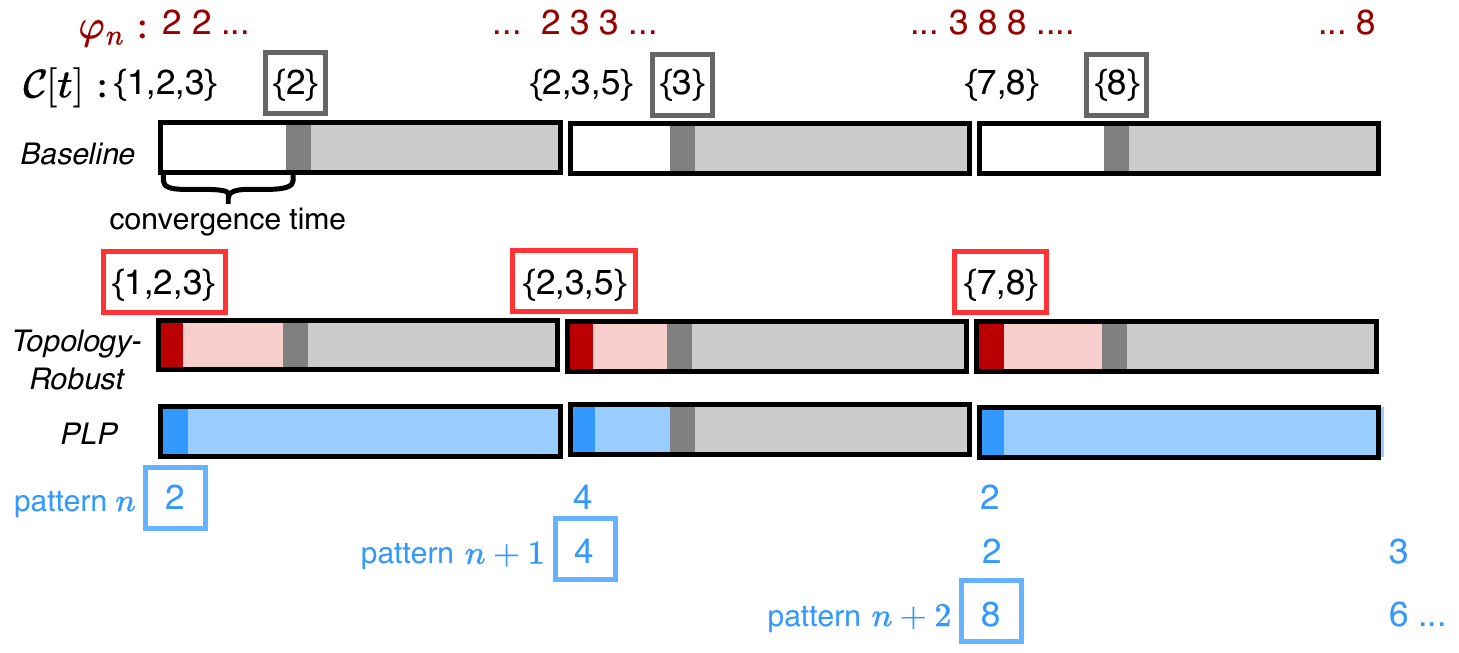}
    \end{center}
    \vspace{-10pt}
    \caption{The time-varying control law for each of the three versions of the controller architecture, designed based on the estimated mode $\hat{\varphi}_{n}^{(t)}$ and the consistent set $\Ccal[t]$.
    Each horizontal bar represents a time duration of length $\Delta T$.
    \authorrevised{
    The baseline uses the previous law until the consistent set converges to a singleton set (white sub-bars).}
    Topology-Robust is able to control multiple modes simultaneously, so it uses a robust law (red sub-bars) until the convergence.
    PLP (future horizon $L{\,=\,}3$) uses the law corresponding to the predicted next mode (blue sub-bars) until convergence; note that when the mode in the converged consistent set is equivalent to the predicted next mode, the control policy need not be changed.
    }
    \vspace{-10pt}
    \label{fig:tradeoff_plp}
\end{figure}

\authorrevised{
The overall control objective is to minimize the mode-dependent quadratic cost function~\eqn{quadratic_cost} subject to constraints imposed by various implementations of SLS from Sec.~\ref{subsec:sls}.
Namely,} we consider three versions of the controller architecture~\fig{two_part_hierarchical_flow}; a visual distinction among the three is shown in~\fig{tradeoff_plp}.

\begin{itemize}[leftmargin=*]
\setlength\itemsep{0em}
\item \textbf{Baseline} [\textit{First row of~\fig{tradeoff_plp}}]: here,~\fig{two_part_hierarchical_flow} is implemented only using Mode Process ID; 
\authorrevised{both PLP and MPC are not used.
The Control Law Design component is implemented with the basic SLS approach from Sec.~\ref{subsec:sls}.
We minimize the cost~\eqn{quadratic_cost} subject to the achievability constraint described by~\eqn{achievable_sls} and the locality constraint described with the sets $\{\Lcal_{i,h}\}_{i\in\Vcal}$.
Because the topology changes over time and basic SLS is not designed for time-varying topologies, this requires the optimization to be solved multiple times.
}

\item \textbf{Topology-Robust} [\textit{Second row of~\fig{tradeoff_plp}}]: we have the same architecture as above, but SLS is replaced with the method of~\cite{han20l4dc}, an extension of SLS to network dynamics under time-varying topological changes.
A single common control law $\boldsymbol{\Phi}^{(i,t)}$ is designed for all consistent modes in $\Ccal[t]$, and this common law is used until time $t^{*}{\,>\,}t$ when $\abs{\Ccal[t^*]}{\,=\,}1$, after which standard SLS is used.

\item \textbf{PLP} [\textit{Third row of~\fig{tradeoff_plp}}]: we combine the original architecture proposed by~\fig{two_part_hierarchical_flow} with the extended SLS approach described in Sec.~\ref{subsec:sls}.
\authorrevised{We minimize the cost~\eqn{quadratic_cost} subject to the data-driven achievability constraint described by~\eqn{achievable_data_sls} and the locality constraint described with the sets $\{\Lcal_{i,h}\}_{i\in\Vcal}$.}
Given pattern collection $\Psi[t]$ at time $t{\,\in\,}\Nbb$ and mode-index $n{\,\triangleq\,}N[t]$, if $\boldsymbol{\psi}{\,\triangleq\,}(\psi_1,\cdots,\psi_L){\,\in\,}\Psi[t]$ is expected to occur at mode-index $n{\,+\,}\Ebb[\hat{\tau}_n^{(t)}]{\,\in\,}\Nbb$, the control law for node $i{\,\in\,}\Vcal$ is scheduled to be $\boldsymbol{\Phi}_{m}^{(i,s)}$, where $m{\,=\,}\psi_1$ and $s{\,\in\,}[T_{n+\Ebb[\hat{\tau}_n^{(t)}]},t^{*})$, where $T_n$ is defined in~\assum{timescale} and $t^{*}$ is the time after $T_{n+\Ebb[\hat{\tau}_n^{(t)}]}$ when $\abs{\Ccal[t^*]}{\,=\,}1$.
For times $s{\,\in\,}[t, T_{n+\Ebb[\hat{\tau}_n^{(t)}]})$ where a prediction is not available, we revert to the baseline controller. 
\end{itemize}

The three architectures are each tested on two specific network systems of the form given in~\eqn{power_plant_dynamics}.
For both systems, the specific $A$ and $B$ matrices in~\eqn{power_plant_dynamics} are the linearized discrete-time power grid dynamics given in Sec. 5 of~\cite{han20l4dc}, which we do not repeat here for the sake of brevity.

\begin{figure}
    \begin{center}
        \includegraphics[width=0.63\columnwidth]{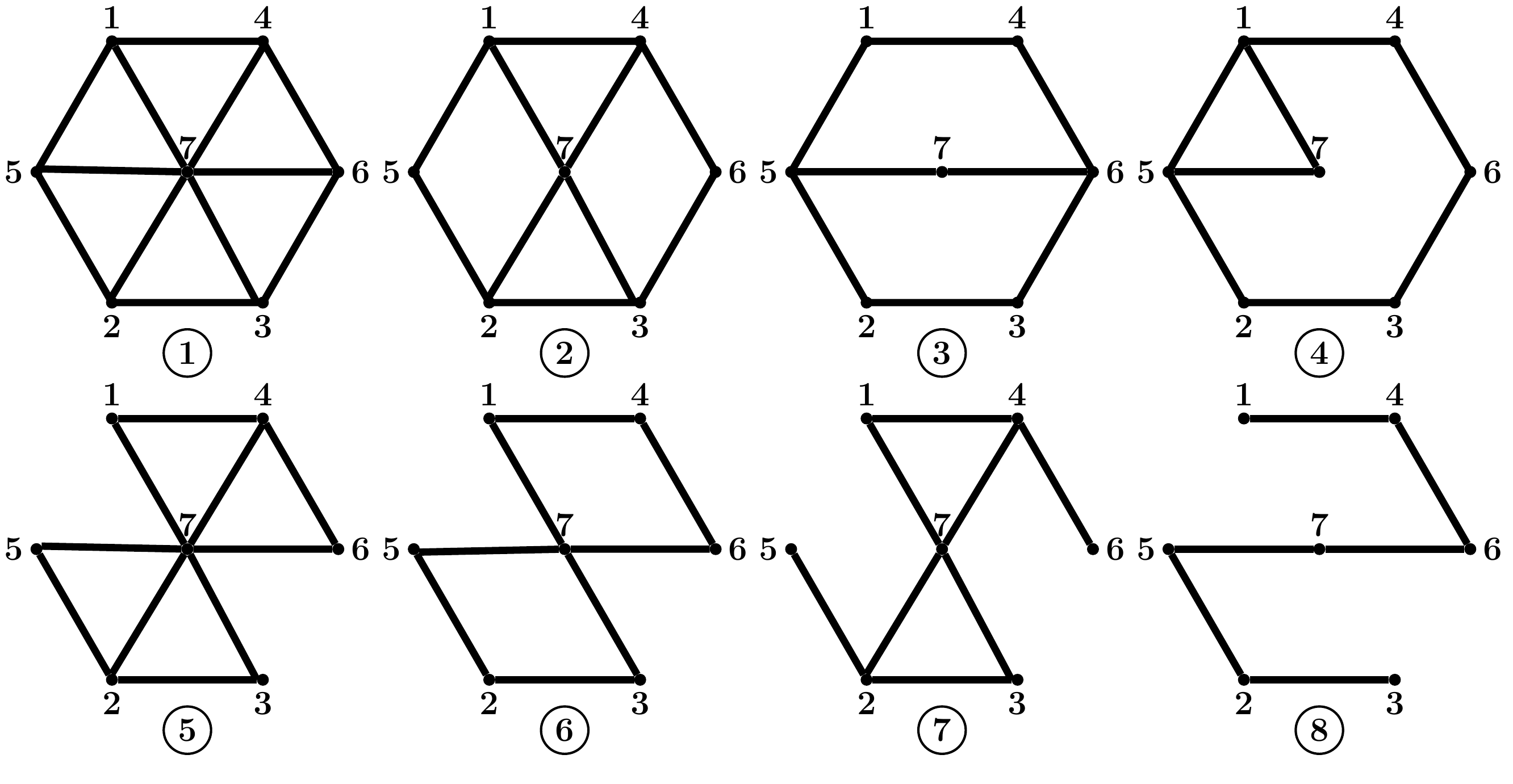}
        \includegraphics[width=0.33\columnwidth]{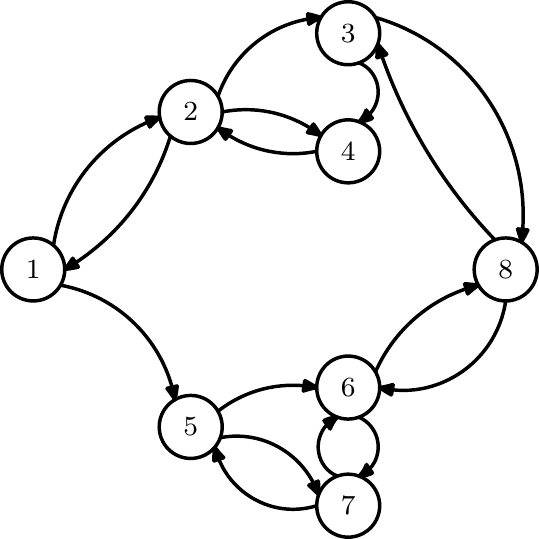}
    \end{center}
    \vspace{-10pt}
    \caption{[Left] The different possible topologies of the Hexagon System.
    [Right] The underlying Markov chain for topology transitions.}
    \label{fig:hexagon_mc}
\end{figure}

\begin{figure}
    \begin{center}
        \includegraphics[width=0.7\columnwidth]{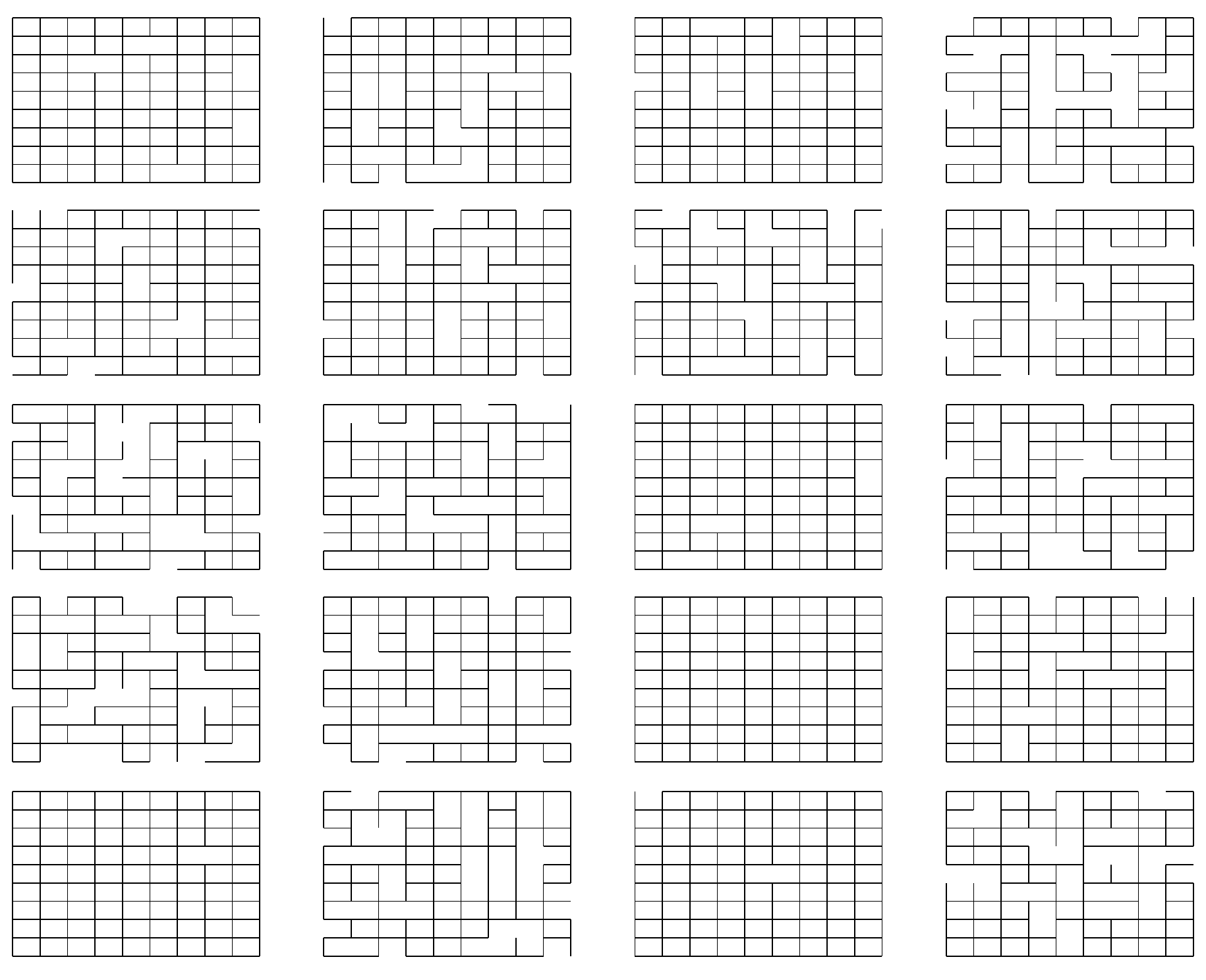}
    \end{center}
    \caption{The different possible topologies of the $10{\,\times\,}10$ Rectangular Grid System.}
    \vspace{-10pt}
    \label{fig:grid_mc}
\end{figure}

\begin{itemize}[leftmargin=*]
\setlength\itemsep{0em}
\item \textbf{(Small-Scale) Hexagon System}: the network system~\eqn{power_plant_dynamics} consists of a hexagonal arrangement of $n_s{\,=\,}7$ nodes and $M{\,=\,}8$ topologies (see~\fig{hexagon_mc}).
When PLP is included, the collection of patterns $\Psi[t]$ is constructed with equality in~\eqn{mpc_pattern_collection}; hence,~\prob{pattern_occurrence} become easy to solve -- every ending string in $\Scal$ is an initial-ending string,
$\Ebb[\hat{\tau}_n^{(t)}]{\,=\,}L$ for each $t{\,\in\,}\Nbb$, $n{\,\triangleq\,}N[t]$, and determining $\text{argmax}_k\{\hat{q}_k^{(t)}\}$ reduces to a maximum likelihood problem.

\item \textbf{(Large-Scale) Rectangular Grid System}: the network system~\eqn{power_plant_dynamics} consists of a $10\times 10$ rectangular grid arrangement of $n_s{\,=\,}100$ nodes and $M{\,=\,}20$ topologies (see~\fig{grid_mc}).
The true TPM is a $M\times M$ stochastic matrix with no self-transitions.
When PLP is included, the collection $\Psi[t]$ is constructed with strict subset in~\eqn{mpc_pattern_collection}, which means the formulas from Theorems~\ref{thm:expected_tau} and~\ref{thm:first_occurrence} must be used to solve~\prob{pattern_occurrence}.
\end{itemize}

\authorrevised{
Even though the Control Law Design component of all three architectures is localized and distributed by the nature of SLS, we initially assume Mode Process ID and PLP are centralized.
This is reasonable under~\assum{timescale}, which imposes that communications among subsystems are much faster compared to the switching of the topologies.
This is often the case in fault-tolerance for large-scale network applications such as the power grid and the internet, where faults are expected to occur rarely.
Furthermore, in Sec.~\ref{subsec:localized}, we introduce the implementation of localized, distributed Mode Process ID and PLP.
For simplification of terminology in this section only, we overload the terminology ``PLP'' to refer to both the controller with PLP (third row of~\fig{tradeoff_plp}) and a component of the controller architecture in~\fig{two_part_hierarchical_flow} that leverages other algorithms, with the understanding that PLP truly refers to the latter.}

\vspace{-5pt}
\subsection{Tradeoff Comparison Results}\label{subsec:tradeoff_plp}
\begin{figure}
    \begin{center}
        \includegraphics[width=\columnwidth]{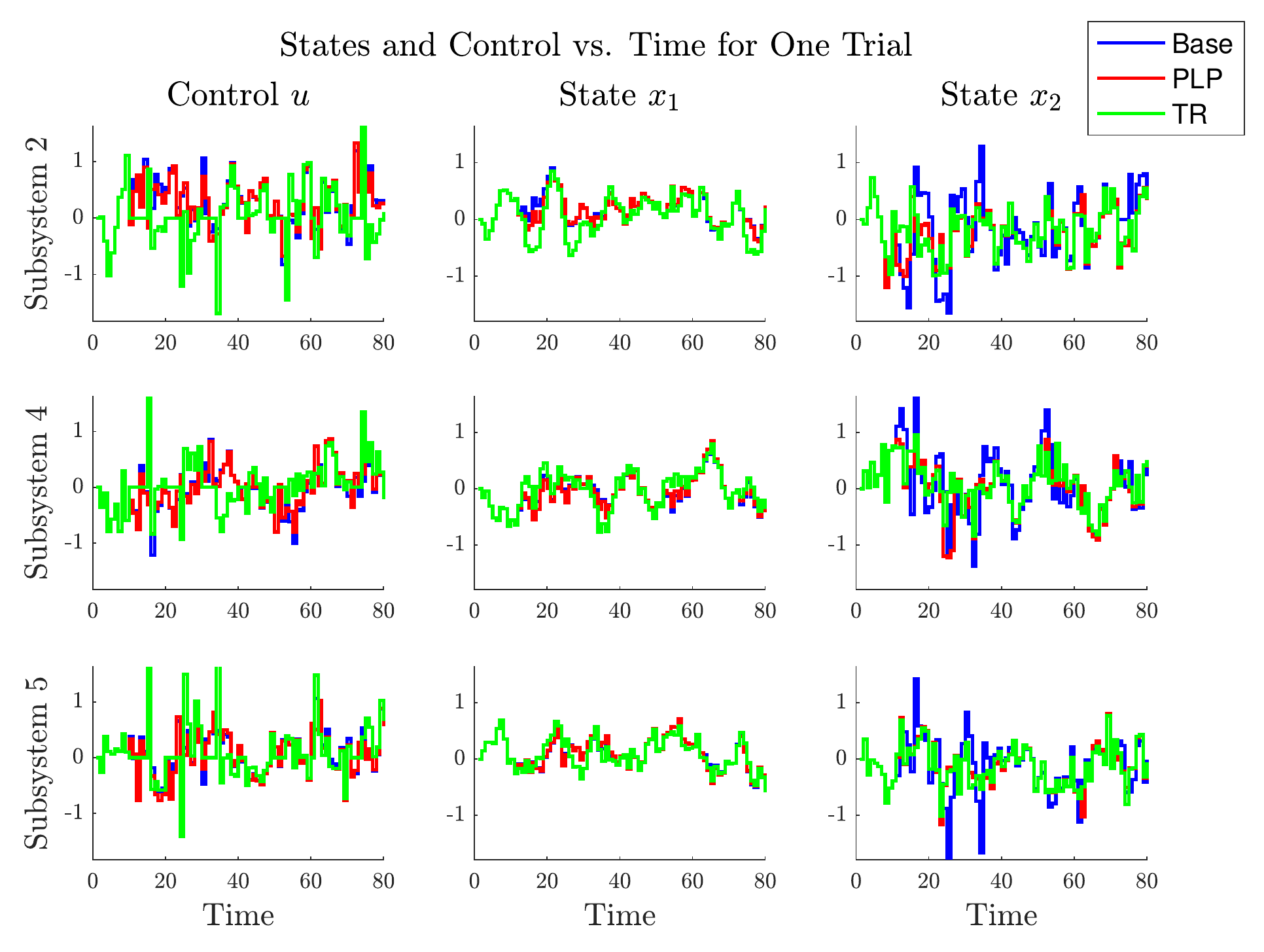}
    \end{center}
    \vspace{-10pt}\caption{\authorrevised{States and control versus time for one Monte-Carlo trial in the hexagon system.
    We abbreviate the baseline controller as `Base', and Topology-Robust as `TR'.}
    }
    \label{fig:states_ctrl}
\end{figure}

\begin{figure}
    \begin{center}
        \includegraphics[width=0.92\columnwidth]{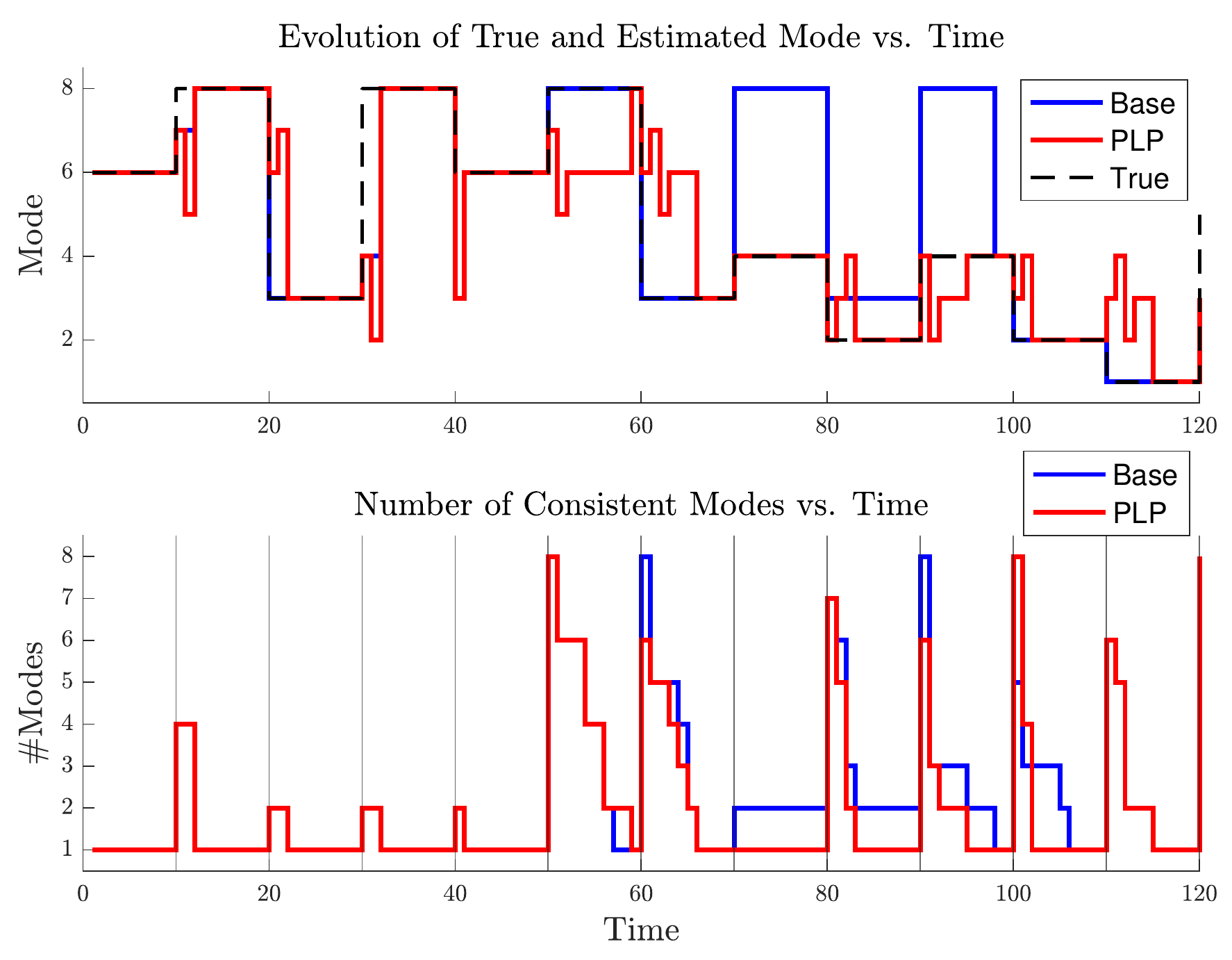}
    \end{center}
    \vspace{-10pt}\caption{\authorrevised{Modes versus time for one Monte-Carlo trial of the hexagon system.
    In the bottom subfigure, black vertical lines indicate intervals of length $\Delta T$.
    }}
    \vspace{-10pt}\label{fig:modes_vs_time}
\end{figure}

Each simulation is run by applying one of the three controller architectures to one of the two network systems.
\authorrevised{
We run a total of $20$ Monte-Carlo experiment trials and each trial is run for $T_{\text{sim}}{\,=\,}400$ timesteps with $\Delta T{\,=\,}10$.
The PLP architecture also uses a future horizon of $L{\,=\,}3$.
A sample trajectory of the states and control versus time for all three architectures is shown in~\fig{states_ctrl} for the hexagon system; we reduce the time horizon to $80$ timesteps for this figure only so that there is better clarity in distinguishing the lines.
Because the objective is the reject external disturbances, the state values waver around the zero line.
Moreover, under Topology-Robust, the state has the smallest oscillations around zero (green), followed by PLP (red), and finally the baseline (blue).
A sample evolution of the consistent set narrowing approach applied for Mode Process ID is also shown in~\fig{modes_vs_time} for the baseline and PLP architectures; again, we plot for a shorter horizon of time (120 timesteps) for easier visibility.
The PLP architecture manages to successfully narrow the consistent set down to a singleton within the $\Delta T$ time interval more often than the baseline, and consequently also manages to track the true mode more precisely.
}

The comparisons among the different scenarios are performed by evaluating one of the following four performance metrics.
First, to measure the control effort, an LQR-like cost~\eqn{lqr_cost} is averaged over the simulation time $T_{\text{sim}}$.
Second, to measure the disturbance-rejection performance, we consider the time-average error norm~\eqn{error}.
Third, we measure the proportion~\eqn{matching} of the simulation duration in which the matching control law is used to control the current topology.
Here, if the true mode is given by $\varphi_{n}$ at time $t$, we say that the \textit{matching control law} $\{\boldsymbol{\Phi}_{m}^{(i,t)}{\,:\,}i\in\Vcal\}$ is used if $m{\,\triangleq\,}\hat{\varphi}_{n}^{(t)}{\,=\,}\varphi_{n}$.
Fourth, the total runtime is recorded. 
\begin{subequations}\label{eq:sim_metrics}
    \begin{align}
        &\hskip1cm \frac{1}{T_{\text{sim}}}\sum\limits_{t=1}^{T_{\text{sim}}} \xvect[t]^{\top}I_{n_x}\xvect[t] + \uvect[t]^{\top}I_{n_u}\uvect[t]\label{eq:lqr_cost}\\
        &\hskip1cm \frac{1}{T_{\text{sim}}}\sum\limits_{t=1}^{T_{\text{sim}}} \norm{\xvect[t]}_2\label{eq:error}\\
        &\hskip1cm \frac{1}{T_{\text{sim}}}\sum\limits_{t=1}^{T_{\text{sim}}} \mathds{1}\{\hat{\varphi}_{n}^{(t)} = \varphi_{n} \}\label{eq:matching}
    \end{align}
\end{subequations}
where $I_{n_x}, I_{n_u}$ are identity matrices with appropriate dimensions.

\begin{figure}
    \begin{center}
        \includegraphics[width=0.95\columnwidth]{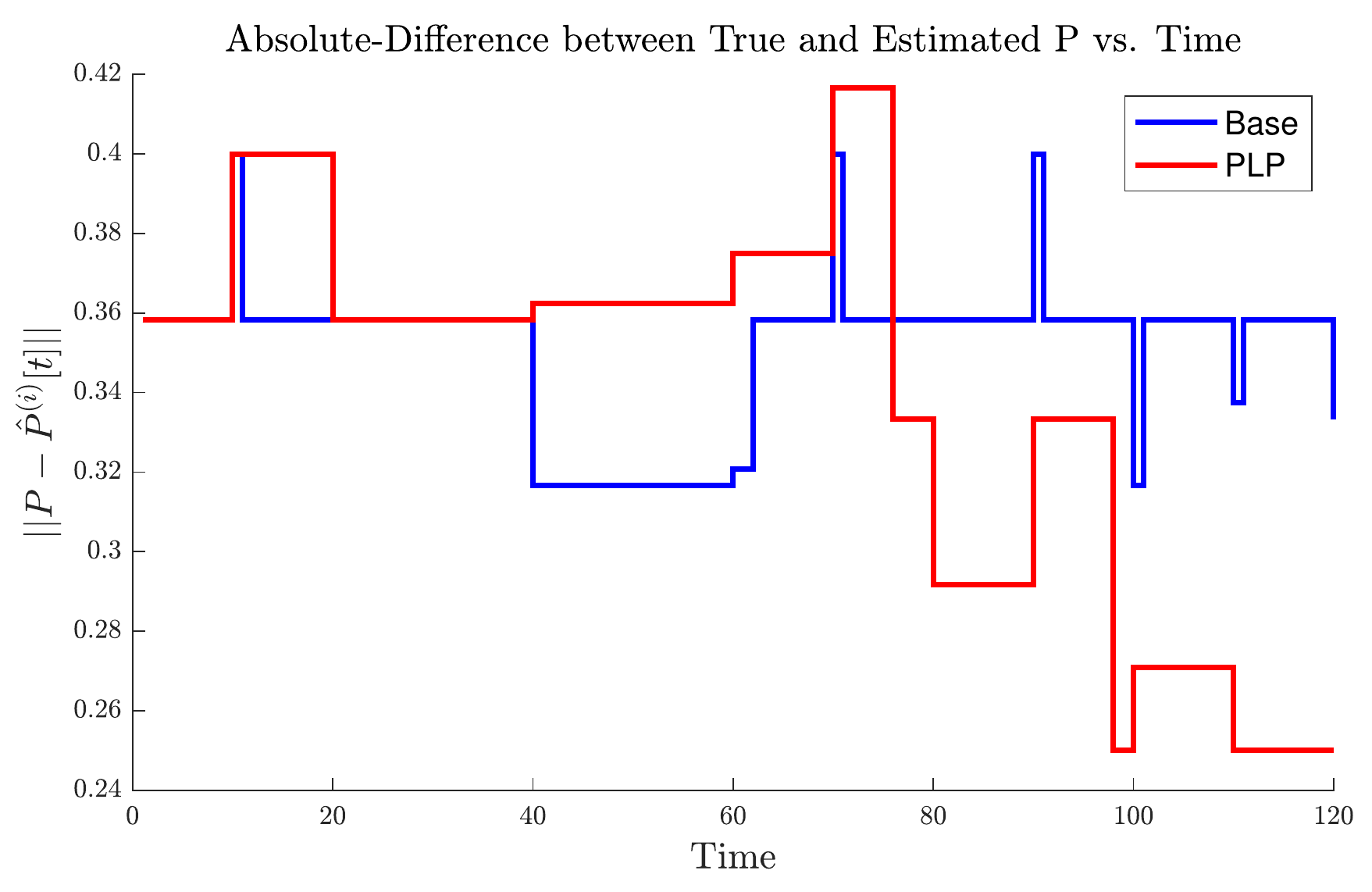}
    \end{center}
    \vspace{-15pt}
    \caption{\authorrevised{Frobenius norm of the difference between the true TPM $P$ and estimated TPM $\hat{P}^{(t)}$ versus time for one Monte-Carlo trial of the hexagon system.}
    }
    \label{fig:mse_tpm}
\end{figure}

The metrics~\eqn{sim_metrics} are further averaged over $20$ Monte-Carlo simulations with varying initial condition $\xvect_0$, noise process $\wvect[t]$, and true realization $\{\varphi_n\}$ of the mode process $\{\xi_n\}$.
\authorrevised{The results are tabulated in~\tab{tradeoff_plp_metrics} with the three architecture names abbreviated: `Base' as the baseline, and `TR' as Topology-Robust.
The proportion of time the matching control law is irrelevant for Topology-Robust because it computes a single law to be used for multiple topologies, hence the `--' entries. 
We also plot a sample evolution of $\norm{P{\,-\,}\hat{P}^{(t)}}$ for one Monte-Carlo trial in~\fig{mse_tpm}, where the norm taken is the Frobenius norm.
Because $\hat{P}^{(0)}$ begins with uniform probabilities in the nonzero positions, there are some variations in the norm difference, but overall, the curve decreases with time, indicating convergence to within a small error ball of the true TPM.
This also allows for the pattern-occurrence quantities to be solved more accurately, which improves the prediction performance of PLP.
As~\tab{tradeoff_plp_metrics} shows, this also enables better controller performance (LQR Cost and Error Norm) of PLP over the other two architectures.}

The values in both sub-rows of the `LQR Cost' row in~\tab{tradeoff_plp_metrics} suggest that the time-average LQR cost of all three controller architectures increase as the scale of the system gets larger.
This is expected because the same values of horizon $H$ and number of hops $h$ (defined in Sec.~\ref{subsec:sls}) were chosen for the SLS implementation of both systems.
In practice, $H$ and $h$ must be adjusted as the scale of the system changes, but for fairer comparison we use the same values for both the hexagon and grid systems.
Furthermore, assuming a small margin of error, Topology-Robust should theoretically stabilize the system better than the baseline at the expense of increased control effort because Topology-Robust uses a single common law is for multiple different modes.
This can be validated empirically by the entries in the `LQR Cost' and `Error Norm' rows,
\authorrevised{
and is also supported by~\fig{states_ctrl}, where the state's oscillations around the zero line are the largest in magnitude with the baseline and the least with Topology-Robust.
}

\bgroup
\def\arraystretch{1.5}
\begin{table}[t]
    \begin{center}
      {\footnotesize
      \begin{tabular}{| c | c | c | c | c |}
        \hline
        Metric / Controller & Base & TR & PLP \\ \hline \hline
        LQR Cost & $36.4537$ & $42.5596$ & $34.7242$\\
        & $445.8137$ & $472.1195$ & $442.1264$ \\ \hline
        Error Norm & $2.2146$ & $1.5546$ & $1.5236$\\
        & $6.1294$ & $5.9453$ & $5.8244$ \\ \hline
        Prop. Match & $0.4304$ & -- & $0.615$\\
        & $0.1533$ & -- & $0.16$ \\ \hline
        Runtime & $11.8314$ & $67.2254$ & $2.2689$\\
        & $101.3741$ & X &  $38.5824$\\\hline
      \end{tabular}
      }
    \end{center}
    \vspace{-10pt}
    \caption{\authorrevised{
    The average performance metrics [row] over $20$ Monte-Carlo simulations of $T_{\text{sim}}{\,=\,}400$ timesteps, for each pair of controller architecture [column].
    In each cell, the top value is recorded for the hexagon system and the bottom is for the grid system.
    For space, we abbreviate `Base' as the baseline controller, and `TR' as Topology-Robust.
    }
    }
    \vspace{-15pt}\label{tab:tradeoff_plp_metrics}
\end{table}
\egroup

More interestingly, the PLP architecture manages to balance the performance metrics better compared to the the other architectures: LQR cost similar to the baseline architecture, error norm similar to the Topology-Robust architecture, and runtime faster than either the baseline or the topology-robust extension.
The improved runtime comes from the PLP component's ability to refrain from recomputing parts of the original SLS optimization by preserving the control inputs of previously-observed topologies and state/control trajectories (see Proposition~\ref{prop:mode_mpc_memory}).
Moreover, the ability of PLP to predict the expected occurrence times of future mode patterns allows for the scheduling of SLS controllers in advance (see Proposition~\ref{prop:mode_mpc_schedule}); as seen in~\fig{tradeoff_plp}, this improves the error norm when Pattern-Learning manages to predict the future mode correctly.
The `Prop Match' row of~\tab{tradeoff_plp_metrics} shows that this is indeed the case: the PLP architecture consistently uses the matching control law more often than the baseline regardless of network system.
This is expected since PLP can be viewed as an additional mode estimation algorithm, and so the estimate $\hat{\varphi}_{n}^{(t)}$ is on average better with PLP than without.
In general, this suggests that appending PLP to a baseline controller that is neither predictive nor robust to time-varying topologies could be used as an alternative to Topology-Robust, especially in complex systems where simultaneous stabilization isn't possible or is expensive.

\authorrevised{We remark that the difference in the construction of the pattern collection $\Psi[t]$} in the hexagon system versus the grid system also has a role in the relationship among the performance metrics, especially in the error-norm performance and the proportion of time the matching control law is used.
Recall that for the hexagon system, $\Psi[t]$ is created by accumulating every feasible mode sequence of length $L$, which implies $\Ebb[\hat{\tau}_n^{(t)}]{\,=\,}L$.
In contrast, for the grid system, a random subset of feasible mode sequences is chosen per time $t$, and so the formulas from Theorems~\ref{thm:expected_tau} and~\ref{thm:first_occurrence} were used to solve~\prob{pattern_occurrence}.
In the PLP column of the `Prop. Match' row, we see the matching control law is used less often in the grid system than the hexagon system, which is expected since $\Ebb[\hat{\tau}_n^{(t)}]{\,\geq\,}L$ for the grid system and predictions for a longer horizon of mode-indices become less accurate.
Thus, increasing the number of patterns in the pattern collection decreases the expected minimum occurrence time, which yields more accurate estimates of future modes.
The Base and PLP columns in the `Error Norm' row suggest that better predictions enable better disturbance-rejection; this implies that PLP will more closely resemble the error norm of the baseline when less patterns are included in $\Psi[t]$.

\subsection{Localized Pattern-Learning and Prediction}\label{subsec:localized}
\tab{tradeoff_plp_metrics} shows that performance deteriorates with larger scale, and this can be attributed to the fact that both Pattern-Learning and Mode Process ID are implemented in a centralized fashion, which conflicts with the localized, distributed nature of SLS.
We now briefly discuss an extension of PLP to a localized, distributed implementation of PLP.
\authorrevised{Since the previous section already compared the performance of PLP to those of the controllers without PLP, we focus our discussion here on how the localized implementation of PLP compares to the centralized version.}

Let current time be $t{\,\in\,}\Nbb$ and $n{\,\triangleq\,}N[t]$. 
Based on information from its own local subsystem~\eqn{power_plant_dynamics}, each node $i{\,\in\,}\Vcal$ stores and updates three objects: a) its own estimates of the current mode $\hat{\varphi}_{n}^{(i,t)}$ and TPM $\hat{P}^{(i,t)}$ (computed via Sec.~\ref{subsec:mode_id}), b) its own estimates of the pattern-occurrence quantities $\Ebb[\hat{\tau}^{(i,t)}]$, $\{\hat{q}^{(i,t)}_k\}_{k=1}^{K}$ (computed via Sec.~\ref{subsec:pattern_learning} and~\ref{subsec:theorems}), and c) its own pattern collection $\Psi^{(i)}[t]$ and pattern-to-control law table $\Ucal^{(i)}$ (see Sec.~\ref{subsec:mode_control}).
Each node $i{\,\in\,}\Vcal$ employs the consistent set narrowing approach of~\eqn{consistent_set} to update its own set $\Ccal^{(i)}[t]$ of consistent topologies over time $t$.
\authorrevised{
Each subsystem $i{\,\in\,}\Vcal$ then extracts $\hat{\varphi}_n^{(i,t)}$, $\Ccal^{(i)}[t]$, and estimates $\hat{P}^{(i,t)}$ by empirically counting the proportion of transitions across the entire estimated past history $\hat{\varphi}_{0:n}^{(i,t)}$.
For the TPMs, we also implement consensus averaging of the estimates to neighboring subsystems that are one link away, similar to the method of Sec. 4 in~\cite{han20l4dc}.}
Overall, the key distinction is that we add an additional enumeration $i{\,\in\,}\Vcal$ to the usual sets, tables, and estimated quantities
\authorrevised{
from Sec.~\ref{sec:mode_id} and Sec.~\ref{sec:pattern_learning}
}
to emphasize that each subsystem maintains local estimates of everything.

\begin{figure}
    \begin{center}
        \includegraphics[width=0.94\columnwidth]{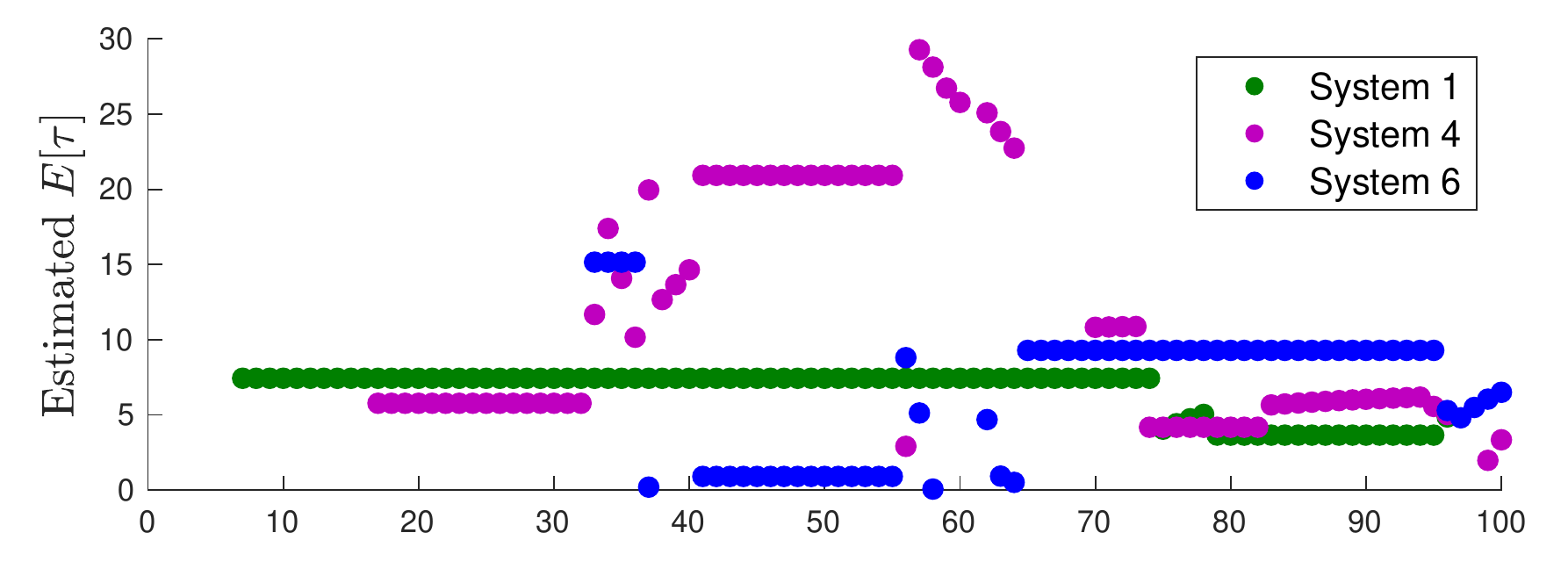} 
        \vskip.1cm
        \includegraphics[width=0.94\columnwidth]{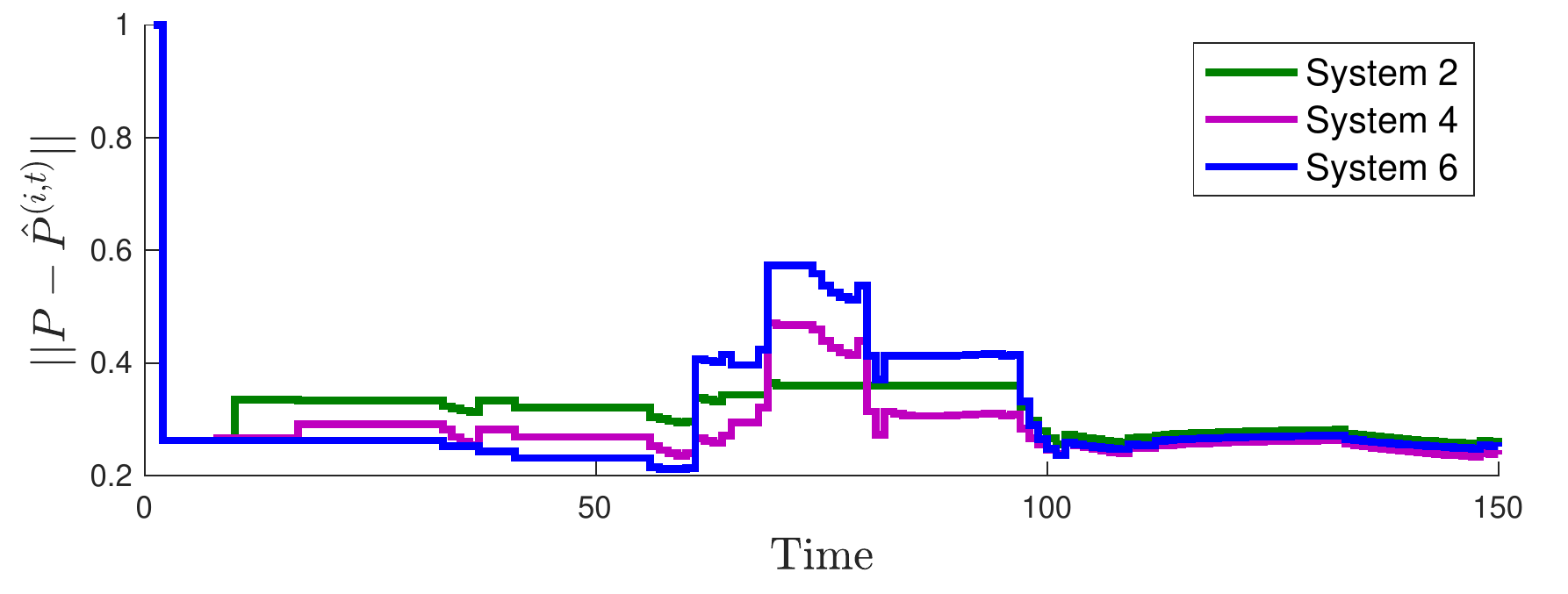}
    \end{center}
    \vspace{-10pt}
    \caption{[Top] The evolution of $\Ebb[\hat{\tau}^{(i,t)}]$ over time for subsystems $i{\,\in\,}\{1,4,6\}$.
    [Bottom] The evolution over time of \authorrevised{the Frobenuis norm of the difference} between $P$ and $\hat{P}^{(i,t)}$ for subsystems $i{\,\in\,}\{2,4,6\}$.
    }
    \vspace{-10pt}
    \label{fig:tpm_taus}
\end{figure}

The estimated pattern-occurrence quantities for this localized extension of PLP applied to the hexagon system are shown in~\fig{tpm_taus}. To demonstrate the evolution of the pattern-occurrence quantities over time, 
\authorrevised{
each subsystem $i$'s pattern collection $\Psi^{(i)}[t]$ is chosen to contain more than half of the full combinatorial set of feasible length-$L$ mode sequences initially considered in Sec.~\ref{subsec:tradeoff_plp}, such that the true value of $\Ebb[\hat{\tau}^{(i,t)}]$ is $5.83328$ via Theorem~\ref{thm:expected_tau}.}
The evolution of the estimated minimum occurrence time $\Ebb[\hat{\tau}^{(i,t)}]$ over $t$ is shown at the top, while the Frobenius norm difference $\|P - \hat{P}^{(i,t)}\|$ of the TPM estimate is shown at the bottom.
We use varying groups of subsystems for these figures in order to demonstrate the locality property.

\authorrevised{Note in~\fig{tpm_taus}, that as time increases, the estimates $\Ebb[\hat{\tau}^{(i,t)}]$ of tend to converge towards the true value $5.83328$ as more of the TPM gets learned.}
The piecewise nature arises because the pattern collection $\Psi^{(i)}[t]$ may change over time, which in turn changes each subsystem's estimate of the expected minimum occurrence time.
At the bottom of~\fig{tpm_taus}, the matrix norm difference between the true and estimated TPMs for each of the three subsystems decreases over time, which is expected as each subsystem gathers more data to learn the true transition probabilities of $P$.
\authorrevised{
Compared to the centralized TPM estimate evolution over time (\fig{mse_tpm}) there is more rapid variation in each subsystem's estimate in the bottom figure of~\fig{tpm_taus}; this could be attributed to the consensus averaging among the subsystems.
Viewing topologies at a local level can make the modes look similar to one another, and so a localized implementation of consistent set narrowing may perform worse than the centralized implementation.
This is a well-known tradeoff between centralized and distributed control: for more efficient computation, we are trading performance optimality.
}
\vspace{-10pt}


\section{Conclusion}\label{sec:conclusion}
\vspace{-10pt}
\authorrevised{\textit{Pattern-learning for prediction (PLP)} learns patterns in the behavior of stochastic uncertain systems to make controller design efficient by memorizing patterns to prevent the re-computation of the control laws associated with previously-occurred patterns (see Proposition~\ref{prop:mode_mpc_memory}) and by scheduling of control laws associated with patterns that may occur in the future (see Proposition~\ref{prop:mode_mpc_schedule}).
In this paper, we aimed to demonstrate the advantages of including PLP in an otherwise straightforward controller architecture (which borrows techniques from system identification and predictive control) for a class of linear MJS whose underlying mode-switching dynamics are unknown; here, the aforementioned patterns are recurrent finite-length sequences of modes which arise in the MJS. 
}
Our controller architecture consists of three parts.
First, Mode Process ID (Sec.~\ref{subsec:mode_id} and~\ref{sec:mode_id}) identifies the unknown statistics of the mode process.
Second, \authorrevised{PLP (Sec.~\ref{subsec:pattern_learning} and~\ref{sec:pattern_learning}) uses the estimated statistics of the mode process to compute the pattern-occurrence quantities from~\prob{pattern_occurrence}:} the expected minimum occurrence time of any pattern from a user-defined pattern collection, and the probability of a pattern being the first to occur among the collection.
\authorrevised{
The computation of the pattern-occurrence quantities uses martingale methods from the literature with two key extensions that make it more applicable to the real-world: 1) the distribution of the mode process is unknown, and 2) the mode process is not observable; closed-form expressions of the quantities are derived in Theorems~\ref{thm:expected_tau} and~\ref{thm:first_occurrence}.
Third, Control Law Design (Sec.~\ref{subsec:mode_control} and~\ref{sec:mode_control}) computes the optimal control action corresponding to each pattern when it first occurs.
We implement PLP on a fault-tolerant controller of a network with dynamic topology by integrating the pattern-occurrence quantities into MPC and using variations of SLS (Sec.~\ref{subsec:sls}) for the Control Law Design component.}
We provide an empirical comparison study of its performance against a baseline controller and a topology-robust extension of the baseline.
Because PLP can be viewed as an additional mode estimation algorithm, it enables the estimated mode to match the true mode more often, although this is mainly possible for an optimal choice of pattern collection.
Compared to the baseline, PLP is able to achieve better disturbance-rejection at reduced computation time, redundancy, and control cost, which suggests its potential to be used in place of a robust controller for more complex applications where designing for robustness is expensive.
The merit of our work can be summarized as follows: computation-efficient control design for stochastic systems with uncertain dynamics can be performed by learning patterns in the system’s behavior, which eliminates redundancy by storing past patterns into memory and predicting the future occurrence of patterns.
\vspace{-5pt}


\section*{ACKNOWLEDGMENTS}
\vspace{-5pt}
The authors would like to thank John Brader and Benjamen Bycroft of the Aerospace Corporation for their technical inputs.

\bibliographystyle{elsarticle-harv}
{\small
\bibliography{%
bibi%
}}



\end{document}